\documentclass[11pt]{article}
\usepackage{amsmath, amssymb, amsfonts, amsthm, euscript, mathrsfs, epsfig, graphics}

\newcommand{\bb}[1]{\begin{equation}\label{#1}}
\newcommand{\ee}{\end{equation}}
\newcommand{\vc}[1]{{\bf #1 }}
\renewcommand{\vec}{\pmb}
\newcommand{\Sum}[1]{\sum\limits_{#1}}
\newcommand{\Int}[1]{\int\limits_{#1}}
\newcommand{\Prod}[1]{\prod\limits_{#1}}
\newcommand{\const}{\operatorname{const}}
\newcommand{\bchi}{\pmb{\chi}}
\newcommand{\trs}{\operatorname{tr}}
\newcommand{\Tr}{\operatorname{Tr}}

\def\identy{{\mathsurround0pt\mathchoice{\textidenty}{\textidenty}{\scptidenty}{\scptidenty}}}
\def\scptidenty{\setbox0\hbox{$\scriptstyle1$}\bothidenty}
\def\textidenty{\setbox0\hbox{$1$}\bothidenty}
\def\bothidenty{\rlap{\hbox to.97\wd0{\hss\vrule height.06\ht0 width.82\wd0}}
 \copy0\rlap{\kern-.36\wd0\vrule height1.05\ht0 width.05\ht0}\kern.14\wd0}

\newcommand{\tK}{{\widetilde K}}

\newcommand{\cH}{{\cal H}}
\newcommand{\cD}{{\cal D}}

\newcommand{\tr}{\mbox{Tr}}

\renewcommand{\a}{\alpha}
\renewcommand{\b}{\beta}

\newcommand{\e}{\varepsilon}

\newcommand{\bR}{{\mathbb R}}
\newcommand{\bC}{{\mathbb C}}

\newcommand{\bZ}{{\mathbb Z}}

\begin{document}
\title{Magnetism and the Weiss Exchange Field - A Theoretical Analysis Inspired by Recent Experiments}
\author{C. Albert\footnote{supported in part by the Swiss National Foundation},\,\, L. Ferrari\footnote{present address: Collegio Papio, Ascona, Switzerland}\,\,, J. Fr\"ohlich\,\,, B. Schlein\footnote{present address: Department of Mathematics, Stanford University, Stanford CA
94305, USA}\\
Theoretical Physics, ETH-H\"onggerberg, CH-8093 Z\"urich}
\maketitle

\begin{abstract}
The huge spin precession frequency observed in
recent experiments with spin-polarized beams of hot electrons shot
through magnetized films is interpreted as being caused by Zeeman
coupling of the electron spins to the so-called {\em Weiss
exchange field} in the film. A ``Stern-Gerlach experiment'' for
electrons moving through an inhomogeneous exchange field is proposed. The
microscopic origin of exchange interactions and of large
mean exchange fields, leading to different types of
magnetic order, is elucidated. A microscopic derivation of the equations of motion of the Weiss
exchange field is presented. Novel proofs of the
existence of phase transitions in quantum $XY$-models and
antiferromagnets, based on an analysis of the statistical
distribution of the exchange field, are outlined.
\end{abstract}

\section{Introduction}
\setcounter{equation}{0}

Effects of ferromagnetism have been known since antiquity. But a mathematical understanding of the microscopic origin of ferromagnetism has remained somewhat elusive, until today! Pauli paramagnetism, ferro-, ferri- and antiferromagnetism are quantum phenomena connected to the spin of electrons and to Pauli's exclusion principle. The theory of {\em paramagnetism} in (free) electron gases is quite straightforward, \cite{A}.
{\em Antiferromagnetism} is relatively well understood: A mechanism for the generation of antiferromagnetic exchange interactions has been proposed by Anderson \cite{5}, who discovered a close relationship between the half-filled Hubbard model and the Heisenberg antiferromagnet using perturbative methods; (see also \cite{6} for mathematically more compelling and more general variants of Anderson's key observation). It has been proven rigorously by Dyson, Lieb and Simon \cite{4}, using the method of infrared bounds previously discovered in \cite{3}, that the {\em quantum Heisenberg antiferromagnet} with nearest-neighbour exchange couplings exhibits a phase transition accompanied by spontaneous symmetry breaking and the emergence of gapless spin waves, as the temperature is lowered, in {\em three} or {\em more} dimensions. (The Mermin-Wagner theorem says that, in (one and) two dimensions, continuous symmetries cannot be broken spontaneously in models with short-range interactions, \cite{B}.)

Our mathematical understanding of {\em ferromagnetism} is far
less advanced. Some kind of heuristic theory of
ferromagnetism emerged, long ago, in the classic works of
Heisenberg, Bloch, Stoner, Dyson, Landau and Lifshitz, and others
\cite{1}. Various insights have been gained on the basis of some
form of mean-field theory, with small fluctuations around
mean-field theory taken into account within a linear
approximation. This approximation, however, is known to break down in the
vicinity of the critical point of a ferromagnetic
material, where nonlinear fluctuations play a crucial role,
\cite{2}.

In a variety of tight-binding models of itinerant electrons,
ferromagnetic order has been exhibited in the {\em ground state}
(i.e., at zero temperature); see \cite{C},\cite{D},\cite{E}. One
of these models is a fairly natural two-band model in which
ferromagnetism arises from a competition between electron hopping,
Coulomb repulsion and an on-site Hund's rule, \cite{E}. (Hund's
rule says that the spin-tripled state of two electrons occupying
the same site is energetically favoured over the spin-singlet
state. It should be emphasized, however, that a mathematically
rigorous derivation of Hund's rule in atomic physics from first
principles has {\em not} been accomplished, so far.) None of the
results in \cite{C},\cite{D},\cite{E} comes close to providing
some understanding of ferromagnetic order and of an order-disorder
phase transition at {\em positive} temperature. It is not known
how to derive, with mathematical precision, an effective
Hamiltonian with explicit ferromagnetic exchange couplings from
the microscopic Schr\"odinger equation, or a tight-binding
approximation thereof, of ferromagnetic materials. But even if we
resort to a phenomenological description of such materials in
terms of models where ferromagnetic exchange couplings have been
put in {\em by hand} we face the problem that we are unable to
exhibit ferromagnetic order at low enough temperature and to
establish an order-disorder phase transition in three or more
dimensions. {\em No mathematically rigorous proof} of the phase
transition in the {\em quantum Heisenberg ferromagnet} is known,
to date! (Such a result has, however, been established for {\em
classical} Heisenberg models in \cite{3}.)

Ab-initio quantum Monte Carlo simulations of models of quantum
ferromagnets are plagued by the well known ``sign- (or
complex-phase) problem''.

Thus, until now, there are neither substantial mathematically rigorous results on, nor are there reliable ab-initio numerical simulations of, realistic models of ferromagnetic metals, such as Ni, Co or Fe! Given that ferromagnetism is among the most striking {\em macroscopic} manifestations - apparent, e.g., in the needle of a compass - of the {\em quantum-mechanical} nature of matter, this is clearly a desolate state of affairs.

In the present paper we shall not remedy this unsatisfactory situation. However, first, we attempt to draw renewed attention to it, and, second, we outline a formalism and some fairly elementary analytical observations of which we hope that they will ultimately lead to a better, mathematically rather precise understanding of ferromagnetism.
Were it not known already, our analysis and the one in \cite{E} would make clear that ferromagnetism is a non-perturbative phenomenon involving strong correlations and gapless modes. To understand it mathematically will most probably necessitate a full-fledged {\em multi-scale} (renormalization group) {\em analysis}. The formalism presented in this paper and our calculations are intended to provide a convenient starting point for such an analysis.

Analytical work on ferromagnetism may seem to be rather
unfashionable. However, there are recent developments, such as
spintronics, fast magnetic devices, etc. that may make work like
ours appear worthwhile. Our own motivation for the work that led
to this paper actually originated in studying recent experiments
with beams of spin-polarized, hot electrons shot through
ferromagnetically ordered films consisting of Ni, Co or Fe that
were carried out in the group of H.C. Siegmann at ETH; see
\cite{7},\cite{8}. Back in 1998, it became clear to one of us that
the concept of the ``{\em Weiss exchange field}'' (see
\cite{9},\cite{13}) would play a useful role in a theoretical
interpretation of the experimental results reported in
\cite{7},\cite{8}. More generally, the Weiss exchange field appears to
offer a key to a systematic study of phase transitions in magnetic
materials, magnetic order, spin precession and magnon dynamics. In this paper, we focus on elucidating the microscopic origin of the Weiss
exchange field, the role it plays in the theory of magnetism, and
its dynamics.

Our paper is organized as follows.

In Section 2, we describe the experiments reported in \cite{7},\cite{8} and sketch a phenomenological interpretation, based on scattering theory, of the results found in these experiments, merely adding some conceptual remarks to the discussion of our experimental colleagues and describing the role played by the Weiss exchange field. We also propose some further experiments, in particular a Stern-Gerlach experiment for electrons traversing an inhomogeneous exchange field.

A mathematically precise analysis of the scattering of electrons (or neutrons, photons, ...) at dynamical targets, such as magnetic films, metallic solids, liquid droplets, ..., will appear elsewhere; (some first results appear in \cite{F}).

In Section 3, we reformulate one-band $t-J$ and Hubbard models in terms of a dynamical Weiss exchange field. For this purpose, the standard imaginary-time functional integral formalism for the analysis of thermal equilibrium states of quantum many-body systems is recalled.
The {\em Weiss exchange field} is seen to be a Lagrange-multiplier field in a Hubbard-Stratonovich transformation of the original functional integral that renders the action functional quadratic in the Grassmann variables describing the electronic degrees of freedom; see e.g. \cite{13},\cite{11}. The effective field theory of the Weiss exchange field is obtained after integrating over those Grassmann variables.
The effective (imaginary-time) action functional of the exchange field and identities for Green functions of spin operators are derived.

In Section 4, we determine the leading terms of the effective action of the Weiss exchange field, ${\bf W}$, in the approximation where fluctuations of the {\em length} of the exchange field are neglected.
For this purpose, we derive the ferro- and antiferromagnetic mean-field equations from the exact effective action of the exchange field.
By solving these equations we determine the most likely length, $W_0$, of the exchange field. From that point on, the length of the exchange field is frozen to be $|\vc W|\equiv W_0$.

We then consider a one-band Hubbard model with a half-filled band and find that, in this situation, the effective action of $\vc W$ is the one of a nonlinear $\sigma$-model with a minimum that favours N\'eel order. This result is found on the basis of controlled perturbative calculations and goes beyond linear stability analysis of the antiferromagnetic mean-field solution, (which has been presented, e.g., in \cite{13}).
It represents a functional-integral version of Anderson's basic observations \cite{5}.

We then turn to ferromagnetically ordered mean field solutions and show that $x$-independent fluctuations are not a source of instability of such a solution.
Then we consider a one-band Hubbard model with a weakly filled, fairly flat band. In this situation one expects that ferromagnetism prevails. Indeed, we find that, at low temperatures, the ferromagnetic mean-field equation has a non-trivial solution, and that this solution belongs to a {\em quadratically stable} critical point of the effective action of {\vc W}. This conclusion is the result of somewhat subtle calculations involving processes close to the Fermi surface, which make the dominant contribution (but would lead to small-energy denominators in a purely perturbative analysis). Details will appear in \cite{G}.
Our calculations support the idea that the one-band Hubbard model with a weakly filled, fairly flat band describes coexistence of metallic behaviour with ferromagnetic order, at sufficiently low temperatures. A similar conclusion was reached, tentatively, in \cite{E} for some two-band Hund-Hubbard models. The methods of the present paper also apply to the model discussed in \cite{E}; see \cite{G}.

In the next to last subsection of Section 4, we exhibit a universal
Wess-Zumino term in the effective action of the Weiss exchange
field $\vc W$ and calculate its coefficient, which is purely
imaginary. The Wess-Zumino term is ``irrelevant'' for
antiferromagnets, but plays a {\em crucial} role in the dynamics
of magnons in {\em ferromagnetically} ordered systems. Repeating
arguments in \cite{17}, we derive the Landau-Lifshitz equations
for magnons in a ferromagnet.

Finally, we draw attention to two well known arguments explaining why there is no magnetic ordering at positive temperature, in one and two dimensions; (but see \cite{B}).

In Section 5, we sketch novel rigorous proofs, based on analyzing the effective field theory of the exchange field, $\vc W$, of the existence of phase transitions and magnetic order at low temperatures in a class of $XY$-models, Heisenberg antiferromagnets and ferromagnets of localized $SO(2n)$- spins, for $n=1,2,\dots$. See \cite{4} for the original results.
Our proof is based on establishing reflection positivity of the effective field theory of the exchange field $\vc W$ and then using the original techniques developed in \cite{3}; (see also \cite{14}, \cite{14II}).

It should be emphasized that the concept of the Weiss exchange field has a number of further, quite exciting applications. We hope to return to these matters in future papers.

Some of the material in this paper has a review character; but
some of it is new. We hope it is fairly easy to read. If it draws
renewed attention to some of the deep technical problems in the
quantum theory of magnetism it has fulfilled its purpose. We
gratefully dedicate this paper, belatedly, to two great colleagues
and friends of the senior author (J.F.): G. Jona-Lasinio, on the
occasion of his seventieth birthday, and H.-C. Siegmann, on the
occasion of his retirement from ETH.

{\it Acknowledgements.} J. Fr\"ohlich thanks S. Riesen and H.-C.
Siegmann for very stimulating discussions of the
experiments in \cite{7}, P. Wiegmann for some crucial advice with the
calculations in Section 4, and the IH\'ES for
hospitality during much of the work on this paper.
We all thank M. Azam for very useful and pleasant discussions on the uses of the exchange field.

\section{Real and gedanken experiments involving the Weiss
exchange field} \label{sec:exp} \setcounter{equation}{0}

\begin{figure}[h]
\begin{center}
\input{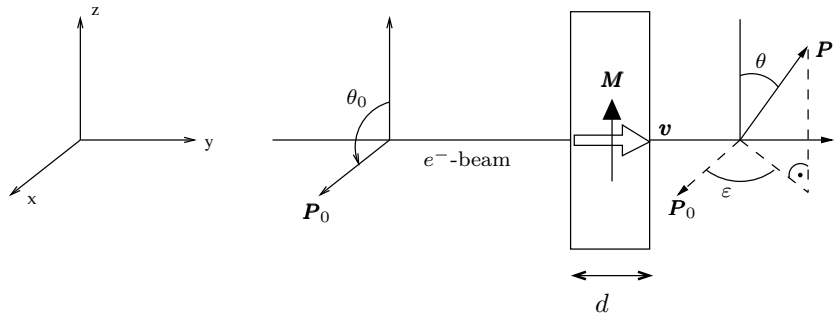}
\end{center}
\caption{Experimental setup}\label{fig:exp1}
\end{figure}

We start this section with a brief description of recent
experiments carried out by Oberli, Burgermeister, Riesen, Weber
and Siegmann at ETH-Z\"urich \cite{7}, \cite{8}. In these experiments, a beam of hot,
spin--polarized electrons is shot through a thin ferromagnetic film (Ni,
Co, or Fe) and the polarization of the outgoing beam is observed.
Their experimental setup is as described in Fig. \protect\ref{fig:exp1}.

The following quantities are measurable:
\begin{itemize}
\item[i)] The thickness, $d$, of the film; $d$ is a few
nanometers. \item[ii)] The average energy, $E$, of an incident
electron; if $E_F$ denotes the Fermi energy of the magnetic film
then $E -E_F$ varies between 4$eV$ and 16$eV$. The group velocity
of the electrons inside the film is denoted by ${\bf v}$; it is
not directly measurable, but it is comparable to
$(2(E+eV)/m_{\text{el}}^*)^{1/2}$, where $eV$ is the average potential energy of an electron and $m_{\text{el}}^*$ its effective mass inside the film.
\item[iii)] The degree, $P_0$, and the direction, ${\bf n}_0$, of
the spin polarization of the incident electron beam; (in Fig.
\ref{fig:exp1}, ${\bf n}_0$ is parallel to the $x$-direction, ${
\bf v}$ to the $y$-direction); the same quantities, $P$ and ${ \bf
n}$, for the outgoing beam. \item[iv)] The direction of the
magnetization, ${ \bf M}$, of the film (in Fig. \ref{fig:exp1}
chosen to be parallel to the $z$-axis); the angles, $\theta_0$ and
$\theta$, between ${ \bf n}_0$ and ${ \bf M}$ and between ${ \bf
n}$ and $\bf M$, respectively ($\theta_0 = \pi /2$, in Fig.
\ref{fig:exp1}). Experimentally, the angle $\theta$ is found to be
considerably smaller than $\theta_0$, i.e., the spins of the
transmitted electrons rotate into the direction of the spontaneous
magnetization ${ \bf M}$ of the film. This is interpreted as being
mainly due to an enhanced absorption of minority-spin electrons,
as compared to majority-spin electrons; see (vi). (It appears that
the contribution of spin flip processes accompanied by magnon
emission into the film -- ``Stoner excitations'' -- to the total
spin rotation is only around 5\%, \cite{7}). \item[v)] The spin
precession angle, $\epsilon$, between the projections of ${ \bf
n}_0$ and of ${ \bf n}$ onto the plane perpendicular to ${ \bf M}$
(the $xy$-plane of Fig. \ref{fig:exp1}); $\epsilon$ is found to be
``large'' (tens of degrees). \item[vi)] Let $I$ be the intensity
of the incident beam, and let $I^+ = I^+ (E)$ and $I^- = I^- (E)$
be the intensities of the outgoing beam of electrons with spin
parallel or antiparallel to ${ \bf M}$, respectively, assuming the
incident beam has intensity $I$ and the spins of its electrons are
parallel to ${ \bf M}$ (${ \bf n}_0$ parallel to ${ \bf M}$, $P_0
\cong 1$), or antiparallel to ${ \bf M}$ (${ \bf n}_0$
anti-parallel to ${ \bf M}$, $P_0 \cong 1$), respectively. Then
$\theta_0 = \theta =0$, or $\theta_0 = \theta =\pi$, respectively,
and $\epsilon =0$. $I^+$ and $I^-$ can be measured and yield the
spin-transmission asymmetry \[ A = \frac{I^+ - I^-}{I^+ + I^-} ;\]
$A$ is found to be positive and large. This is interpreted in
terms of rates of transitions of electrons into unoccupied 3d
states (holes): There are more unoccupied 3d states in the film
with spin antiparallel to ${ \bf M}$ (minority spin) than with
spin parallel to ${ \bf M}$ (majority spin). This explains
qualitatively the experimental results found for $A$ and for
$\theta - \theta_0$ (see (iv)); \cite{7}. \item[vii)] The orbital
deflection angle, $\alpha$, between the directions of the incident
and the transmitted beam (not indicated in Fig. \ref{fig:exp1}).
Experimentally $\alpha$ is found to be negligibly small. This
tells us that the integrated Lorentz force on the electrons
transmitted through the film is tiny. The precession of the spins
of the electrons when they traverse the film can therefore {\it
not} be explained by Zeeman coupling of the spins to the magnetic
field inside the layer. It is mainly due to Zeeman coupling of the
spins to what will be called the {\it Weiss exchange field}. In
iron, the Weiss exchange field causing the observed spin
precession would correspond to a magnetic field of roughly 8000
Tesla (which is gigantic).
\end{itemize}

A theoretical interpretation of the experimental results reported
in \cite{7} can be attempted within the formalism of {\it
scattering theory}. If the luminosity of the incident beam is low
we can consider a single incoming electron. The incoming state is
described as a tensor product of a Pauli spinor,
$\psi_{\text{in}}$, describing the incident electron and a state,
$\xi$, of the film. Typically, $\xi$ is the ground state
(temperature $T=0$) or a thermal equilibrium state ($T >0$) of the
film. The outcoming state, long after the interactions between an
outgoing electron and the film have taken place, is more
complicated and will, in general, exhibit {\it entanglement}
between the electron and the degrees of freedom of the film. If
only measurements far away from the film are performed, as in
\cite{7}, the outgoing state can be described as a density matrix
\begin{equation}
P_{\text{out}} = (\rho_N)_{N=0}^{\infty},
\end{equation}
where $\rho_N$ is a non-negative, trace-class operator on the
Hilbert space of $N$ {\it outgoing electrons} (the incident
electron has knocked $N-1$ electrons out of the film), $N
=2,3,\dots$; $\rho_0$ is a non-negative number, the {\it
absorption probability}, $\rho_1$ is a non-negative trace-class
operator on the Hilbert space
\begin{equation}
\mathcal{H} = L^2 (\bR^3) \otimes \bC^2
\end{equation}
of square-integrable Pauli spinors and describes the (generally
{\it mixed}) state of {\it one} outgoing electron. ``Conservation
of probability'' implies that
\begin{equation}\label{eq:cons}
\rho_0 + \sum_{N=1}^{\infty} \tr \rho_N =1 .
\end{equation}
The state $P_{\text{out}}$ is obtained by taking a partial trace
of the outgoing state of the {\it total system, including} the
film, over the degrees of freedom of the film. This is justified,
because the degrees of freedom of the film are {\it not} observed
in the experiment. If the energy, $E$, of the incident electron is
below (or comparable to) the threshold, $\Sigma^{(2)}$, for
emission of two or more electrons from the film then
\begin{equation}\label{eq:rhoN}
\rho_N = 0 \quad \text{for } N \geq 2.
\end{equation}
In the interpretation of the experimental data provided in
\cite{7}, this is tacitly assumed.

Experimentally, the absorption probability $\rho_0$ and the spin
polarizations ${ \bf P}_0$ and ${ \bf P}$ of the incoming and the
outgoing electron, respectively, are measured. The vectors ${ \bf
P}_0$ and ${ \bf P}$ are given by
\begin{equation}
{ \bf P}_0 \equiv P_0 { \bf n}_0 = \langle \psi_{\text{in}} , {
\pmb \sigma} \psi_{\text{in}} \rangle ,
\end{equation}
and
\begin{equation}
{ \bf P} \equiv P { \bf n} = \frac{1}{ \tr_{\cH} (\rho_1)} \,
\tr_{\cH} (\rho_1 { \pmb \sigma}) ,
\end{equation}
where $\psi_{\text{in}}$ is the wave function of the incident
electron, ${ \pmb \sigma} = (\sigma_x, \sigma_y,\sigma_z)$ is the
vector of Pauli matrices, and $\widetilde \rho_1 := [\tr_{\cH}
(\rho_1)]^{-1} \rho_1$ is the {\it conditional} state of the
outgoing electron, given that it has {\it not} been absorbed in
the film. If (\ref{eq:rhoN}) is assumed to hold then
\begin{equation}
\tr_{\cH} (\rho_1) = 1 -\rho_0
\end{equation}

Since we have assumed that the incoming electron has been prepared
in a pure state,
\begin{equation}
P_0 \equiv |{ \bf P}_0|=1 .
\end{equation}
However, there is, a priori, no reason why $\widetilde \rho_1$
should be a pure state. If it were pure then
\begin{equation}
P = |{ \bf P}|=1 .
\end{equation}

It would be highly interesting to estimate, experimentally, the
amount of entanglement with the film (or {\it decoherence}) in the
state of the outgoing electron by measuring the quotient $P/P_0$.
If $P /P_0 <1$ then $\widetilde \rho_1 = [\tr_{\cH} (\rho_1)]^{-1}
\rho_1$ is {\it not} a pure state, anymore, meaning there is
entanglement with the film. Apparently, $P/P_0$ has not been
measured accurately, yet.

A moment's reflection shows that spin flip processes accompanied
by magnon emission in the film (``Stoner excitations'') lead to
entanglement; while absorption of electrons into unoccupied 3d
states need not be correlated with entanglement of the states of
{\it those} electrons that {\it do} traverse the film. In fact, it
is implicitly assumed in \cite{7} that if Stoner excitations are
neglected then $\widetilde \rho_1$ is close to a pure state (at
least in spin-space). The experimental techniques of \cite{7}
could be used to test this hypothesis.

Next, we express the spin transmission asymmetry $A$ (see (vi)) in
terms of outgoing states. Let $\psi^+_{\text{in}}$ and
$\psi^-_{\text{in}}$ be incoming states with spin polarization ${
\bf P}_0 \equiv { \bf P}_0^+$ parallel to ${ \bf M}$ (majority
spin) and ${ \bf P}_0 \equiv { \bf P}_0^-$ antiparallel to ${ \bf
M}$ (minority spin). Let $P^+_{\text{out}} =
(\rho_N^+)_{N=0}^{\infty}$ and $P_{\text{out}}^- =
(\rho_N^-)_{N=0}^{\infty}$ be outgoing states corresponding to
$\psi_{\text{in}}^+$, $\psi_{\text{in}}^-$, respectively. Then
\begin{equation}\label{eq:Asy}
 A= \frac{\rho_0^- - \rho_0^+}{2 - \rho_0^+ - \rho_0^-},
\end{equation}
as follows from (\ref{eq:cons}). More interesting would be
measurements of
\begin{equation}
{ \bf P}^{\pm} = \tr (\widetilde \rho_1^{\pm} { \pmb \sigma}).
\end{equation}
Clearly ${ \bf P}^{\pm}$ are parallel or anti-parallel to ${ \bf
M}$; but their lengths $P^{\pm} = | { \bf P}^{\pm}|$ ought to be
measured. In \cite{7}, it is tacitly assumed that $P^{\pm} \cong
1$, and that the states $\widetilde \rho_1^{\pm}$ are close to
{\it pure} states; but serious experimental data backing up this
hypothesis appear to be lacking. It is clear that it would be
invalidated if ``Stoner excitations'' played an important role.

In the following, we outline a phenomenological description of the
experiments in \cite{7}, assuming that (\ref{eq:rhoN}) and the
hypothesis just discussed (purity of $\widetilde \rho_1^{\pm}$)
are valid. (A more detailed discussion of the scattering approach
to electron transmission-- and reflection experiments will be
presented elsewhere.)

When an incoming electron enters the film it occupies an empty
state of the film. If the film is crystalline this state belongs
to a band of states; let $\alpha$ be the corresponding band index.
The state of an electron in band $\alpha$ is described by a Pauli
spinor
\begin{equation}
\phi = (\phi^+,\phi^-) ,
\end{equation}
where $\phi^+$ and $\phi^-$ are the components of $\phi$ with spin
parallel or anti-parallel to the magnetization ${ \bf M}$,
respectively. Adopting the approximation of the Peierls
substitution, the Pauli equation for $\phi$ in configuration space
has the form
\begin{equation}\label{eq:cD0}
i \hbar \cD_0 \phi = \mathcal{E}_{\alpha} (-i\hbar { \bf \cD})
\phi,
\end{equation}
where $\mathcal{E}_{\alpha} ({ \bf p})$ is the band function of
band $\alpha$,
\begin{align}
\cD_0 &= \frac{\partial}{\partial t} + i \frac{e}{\hbar} \phi^c +
i \left({ \bf W}_0^c \cdot \frac{{ \pmb \sigma}}{2}\right), \label{eq:cD-0}\\
\cD_j &= \frac{\partial}{\partial x^j} + i \frac{e}{\hbar} A_j + i
\left({ \bf W}_j \cdot \frac{{ \pmb \sigma}}{2}\right).
\end{align}
Here $\phi^c$ denotes a (complex) electrostatic potential, ${ \bf
W}_0^c$ is a (complex) {\it Weiss exchange field}, $(A_j) =
(A_1,A_2,A_3)$ is the electromagnetic vector potential, and $({
\bf W}_j) = ({ \bf W}_1 , { \bf W}_2, { \bf W}_3)$ is an
$SU(2)$-vector potential responsible for {\it spin-orbit
interactions}. As discussed in \cite{9}, (\ref{eq:cD0}) displays
electromagnetic $U(1)$-{\it gauge invariance and} $SU(2)$-{\it
gauge invariance}, i.e., covariance with respect to local
$SU(2)$-rotations in spin space. A number of important
consequences of these gauge symmetries have been pointed out in
\cite{9}.

According to (vii) above, effects of the electromagnetic vector
potential ${ \bf A}$ are apparently negligible; so ${ \bf A}$ is
set to 0. The electrostatic vector potential $\phi^c$ is given,
approximately, by
\begin{equation}
\phi^c = V + i\hat v
\end{equation}
where $eV$ is the surface exit work, and the imaginary part, $\hat v$,
of $\phi^c$ provides a phenomenological description of
spin-independent inelastic absorption processes inside the film.
Velocity-dependent spin-flip processes due to {\it spin-orbit
interactions} appear to play a very minor role in the experiments
reported in \cite{7}; so we may set ${ \bf W}_j$ to 0, $j=1,2,3$.

The Weiss exchange field ${ \bf W}_0^c$ is given by
\begin{equation}\label{eq:W0c}
{ \bf W}_0^c = { \bf W} -i{ \bf w},
\end{equation}
where the real part ${ \bf W}$ describes {\it exchange
interactions} between the incoming electron and the electron
density of the film, and the imaginary part, ${ \bf w}$, yields a
phenomenological description of spin-dependent absorption
processes.

Let $\phi_{\text{in}}$ denote the Pauli spinor describing the
state of an electron when it enters the film at some time $t_0$.
Eq. (\ref{eq:cD0}) can be solved for $\phi =\phi_t$, $t \geq t_0$,
with $\phi_{t_0} = \phi_{\text{in}}$. The solution is explicit if
${ \bf A}=0$, ${ \bf W}_j = 0$, $j=1,2,3$. Let us suppose that the
real part ${ \bf W}$ and the imaginary part ${ \bf w}$ of the
exchange field ${ \bf W}_0^c$ are both anti-parallel to the
magnetization ${ \bf M}$ of the film. Then (\ref{eq:cD0}) leads us
to consider two simple, quasi-one-dimensional scattering problems
in a complex potential well of depth
\begin{equation}
eV + ie\hat v \pm \frac{1}{2} (\Omega -i\omega),
\end{equation}
with $\Omega = |{ \bf W}|$ and $\omega =|{ \bf w}|$, for electrons with
spin parallel to ${ \bf M} (+)$, or anti-parallel to ${ \bf M}
(-)$, respectively. The solution to these scattering problems can
be found in every book on elementary quantum mechanics. For the
purposes of interpreting the results in \cite{7}, a semi-classical
treatment appears to be adequate. The group velocity, ${ \bf
v}_{\pm}$, of an incoming electron wave with energy peaked at $E$
and spin up $(+)$, or down $(-)$, inside the film can be found by
solving the equation
\begin{equation}\label{eq:xip}
\mathcal{E}_{\alpha} ({ \bf p}) = E-eV \mp \frac{1}{2} \Omega
\end{equation}
for ${ \bf p}$ and then setting
\begin{equation}
{ \bf v}_{\pm} = \frac{\partial \mathcal{E}_{\alpha}}{\partial {
\bf p}} ( { \bf p}^{\pm}),
\end{equation}
where ${ \bf p}^{\pm}$ is a solution of (\ref{eq:xip}) chosen such
that ${ \bf v}_{\pm}$ points in the positive $y$-direction. The
sojourn time $\tau_{\pm}$ of the wave inside the film is then
given by
\begin{equation}
\tau_{\pm} = \frac{d}{v_{\pm}}, \quad \text{with} \quad v_{\pm} =
|{ \bf v}_{\pm}|.
\end{equation}
If $\phi_{\text{in}}^{\pm}$ is the state of the electron with spin
up $(+)$, or spin down $(-)$, respectively, when it enters the
film its state $\phi_{\text{out}}^{\pm}$ upon leaving the film is
then given, approximately, by \footnote{Here and henceforth, we
use units such that $\hbar =1$.}
\begin{equation}\label{eq:phiout}
\phi_{\text{out}}^{\pm} \cong \exp \left( -i\tau_{\pm} [ (E-eV)
-ie\hat v \mp \frac{1}{2} (\Omega -i\omega)]\right) \phi_{\text{in}}^{\pm} .
\end{equation}
The presence of inelastic absorption processes implies that $e\hat v
\mp \omega/2 >0$. Setting
\begin{equation}
\exp \left( - \tau_{\pm} [\hat v \mp \omega/2]\right) = c \sqrt{1 \pm A},
\end{equation}
we find that
\begin{equation}\label{eq:phiout2}
\phi^{\pm}_{\text{out}} = c \sqrt{1 \pm A} e^{i(\theta \pm
\epsilon /2)} \phi^{\pm}_{\text{in}},
\end{equation}
where $A$ is the spin transmission asymmetry (see point (iv) and
(\ref{eq:Asy})), $\epsilon$ is the spin precession angle, and
$\theta$ is an (unimportant) spin-independent phase. If $v_+ \cong
v_- = v$ then $\tau_+ \cong \tau_- \cong d/v$, and
(\ref{eq:phiout}) yields
\begin{equation}
\epsilon \cong \tau_{\pm} \Omega \cong (d/v) |{ \bf W}| .
\end{equation}
Thus, measuring $\epsilon$ and $d$ and estimating $v$ yields an
approximate value for the size of the spin precession angular
velocity $\Omega$ and hence of the size of the Weiss exchange
field.

Eqs. (\ref{eq:phiout}) and (\ref{eq:phiout2}) could be mistaken
for equations describing {\it kaon}- or {\it neutrino
oscillations} and are analogous to the equations describing the
{\it Faraday rotation} of light traversing a magnetized medium.

The origin of the Weiss exchange field ${ \bf W}$ is hardly a
mystery: The spin of an electron traversing the film in a band
$\alpha$ apparently experiences (exchange) interactions with the spins
of the occupied states of the film. Since states with spin up and
with spin down are occupied asymmetrically (corresponding to the
fact that ${ \bf M} \neq 0$), the net spin density, ${ \bf S}
(x)$, at a point $x$ in the film is different from zero. The Weiss
exchange field ${ \bf W} = { \bf W} (x)$ is given by
\begin{equation}\label{eq:Wx}
{ \bf W} (x) = - J_{\alpha} { \bf S} (x),
\end{equation}
where $J_{\alpha}$ is the strength of the {\em exchange coupling}
between spins in the $\alpha^{\text{th}}$ band and those in the
occupied band. The theoretical discussion above is based on a {\it
mean field ansatz}: The exchange field ${ \bf W}$ in
(\ref{eq:cD-0}), (\ref{eq:W0c}) is chosen to be
\begin{equation}
{ \bf W} = \langle { \bf W} (x) \rangle = - J_{\alpha} \langle {
\bf S} (x) \rangle.
\end{equation}
(Our conventions are such that $\langle { \bf S} (x) \rangle$ is
parallel to the magnetization ${ \bf M}$.) One of the surprising
implications of the experimental results of \cite{7} is that,
apparently, $J_{\alpha}$ is quite large, even for rather
high-lying bands ($\alpha$), implying that the orbitals of states in such
bands must have substantial overlap with those of states in the
partially occupied, spin-polarized band.

Eq. (\ref{eq:Wx}) makes it clear that ${ \bf W} (x)$ is a {\it
dynamical} field. One of the main purposes of this paper is to
derive the effective quantum dynamics of ${ \bf W} (x)$ within a
Lagrangian functional integral formalism and to sketch what can be
accomplished with this formalism.

Before proceeding with this program, we outline some gedanken
experiments in the spirit of the experiments in \cite{7}.
\begin{itemize}
\item[i)] By applying an external magnetic field to the film
rotating in the $xy$-plane with angular velocity $\omega_0$, the
exchange field ${ \bf W}$ could be made to rotate around the
$z$-axis:
\begin{equation}
{ \bf W} (t) = W \left( \epsilon \cos (\omega_0 t + \delta) ,
\epsilon \sin (\omega_0 t + \delta) , \sqrt{1-\epsilon^2} \right).
\end{equation}
A polarized beam of electrons shot through this film must exhibit
{\it Bloch spin resonance}; but, in this experiment, it would be
due to the rotation of the exchange field. \item[ii)] One may
envisage a {\it Stern Gerlach experiment} for electrons. One would
start by constructing a sandwich of two ferromagnetic metals, I
(e.g. Fe) and II (e.g. Ni), with exchange fields ${ \bf W}_{I}$
and ${ \bf W}_{II}$ of different strength, joined by a transition
region, a mixture of I and II, of width $d_0$. The transition
region would be parallel to the $xy$-plane (see Fig.
\ref{fig:exp2}). One shoots an unpolarized beam of (not very hot)
electrons through the sandwich along the transition region between
I and II, as shown in Fig. \ref{fig:exp2}. One would expect to
detect two beams emerging on the other side of the film in
slightly different directions that are spin-polarized in opposite
directions.

\begin{figure}[h]
\begin{center}
\input{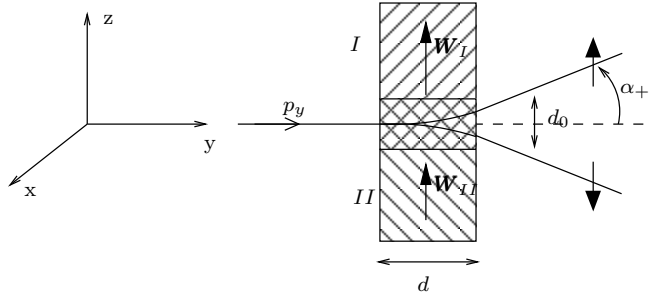}
\end{center}
\caption{Stern-Gerlach Experiment}\label{fig:exp2}
\end{figure}

The force, $f$, in the $z$-direction on an electron
with spin up/down inside the film is given, approximately, by
\begin{equation}
f \cong \pm \frac{|{ \bf W}_I - { \bf W}_{II}|}{2 d_0}.
\end{equation}
It yields a change in the $z$-component of the momentum of the
electron during its passage through the film given by
\begin{equation}
\Delta p_z \cong \pm \frac{|{ \bf W}_I - { \bf W}_{II}|}{2 d_0} \,
\frac{d}{v} ,
\end{equation}
where $v$ is the average group velocity. The deflection angle
$\alpha$ is found from
\begin{equation}
\tan \alpha = \frac{\Delta p_z}{\Delta p_y} .
\end{equation}
Of course, as discussed above, the intensity, $I_{+}$ of the upper
beam can be expected to be much larger than the intensity, $I_-$,
of the lower beam, due to spin-asymmetric absorption inside the
film. In order to achieve adequate focussing, the incident beam
could come from the tip of a scanning tunnelling microscope.
\item[iii)] In an experimental set-up similar to the one above,
one would force an electron current, spin-polarized in the
$z$-direction, through the film. Then a Hall tension in the
$z$-direction should be observed (Hall effect for spin currents;
see \cite{9}).
\end{itemize}


\section{The Weiss exchange field in general one-band ($t-J$ and Hubbard) tight-binding models}
\label{sec:mf}\setcounter{equation}{0}

In this section, we study the origin and dynamics of the Weiss
exchange field in simple tight-binding models. We start by
considering one-band models; but it is straightforward to include
higher (partially occupied or empty) bands relevant for the study
of electron transmission through magnetic films.

Every site, $x$, of a lattice $\Gamma\overset{e.g.}= \mathbb Z^d$, $d=1,2,3,\dots$ carries a four-dimensional state space, $\mathbb C^4$, corresponding to an empty state, a one-electron state with spin up ($+$) or down ($-$), or a state of two electrons in a spin-singlet state. The corresponding basis vectors of $\mathbb C^4$ are denoted by $\{ |0\rangle_x, |+\rangle_x,|-\rangle_x,|+-\rangle_x\}$, $x\in \Gamma$. We introduce electron creation- and annihilation operators, $c_s^\dag(x)$ and $c_s(x)$, respectively, where $s=\pm$, $x\in \Gamma$, satisfying canonical anti-commutation relations
\begin{equation}\label{3.1}
\{c_s^\#(x),c_{s'}^\#(x')\}=0\,,\quad
\{c_s^\dag(x),c_{s'}(x')\}=\delta_{ss'}\delta_{xx'}\,,
\end{equation}
with $c^\#=c$ or $c^\dag$. Then
\begin{equation}\label{3.2}
c_s(x)|0\rangle_x=0\,,\quad \text{for} \quad s=\pm\,, x\in\Gamma
\end{equation}
and
\bb{3.3}
|s\rangle_x=c_s^\dag(x)|0\rangle_x\,,\quad \text{etc.}
\ee
Number operators are given by
\bb{3.4}
n_s(x)=c_s^\dag(x)c_s(x)\,,\quad n(x)=n_+(x)+n_-(x)\,,
\ee
where $n_s(x)$ measures the number of electrons at site $x$ with spin $s$.
By (\ref{3.1}), $n_s(x)$ has eigenvalues $0$ and $1$, while $n(x)$ has eigenvalues $0$, $1$ and $2$. The spin operator, $\vc{S}(x)$, at site $x$ is given by
\bb{3.5}
\vc{S}(x)=\frac{1}{2}\Sum{s,s'}c_s^\dag(x){\pmb \sigma}_{ss'}c_{s'}(x)
\ee
where ${\pmb \sigma}_{ss'}$ is the $(ss')$-matrix element of the vector of Pauli matrices ${\pmb \sigma}=(\sigma_x,\sigma_y,\sigma_z)$.
The operators $\vc S(x)$ vanish on $|0\rangle_x$ and $|+-\rangle_x$, while, for any linear combination $\phi(x)$ of $|+\rangle_x$ and  $|-\rangle_x$,
\bb{3.6}
\vc S(x)\cdot \vc S(x) \phi(x) = \frac{3}{4} \phi(x)\,.
\ee

The dynamics of a gas of electrons moving on the lattice $\Gamma$ is assumed to be generated by a $t-J$ {\em model Hamiltonian} of the form
\bb{3.7}
H=T+U+E_{\neq}\,,
\ee
where the kinetic energy operator, $T$ (the hopping term), is given by
\bb{3.8}
T=\Sum{x{\neq} y}\hat t(x-y)\Sum{s}c_s^\dag(x)c_s(y)\,,
\ee
\bb{3.9}
U=U_0\Sum{x}n_+(x)n_-(x)
\ee
describes an on-site (Hubbard) repulsion, and
\bb{3.10}
E_{{\neq}}=-\frac{1}{2}\Sum{x,y}J_{\neq} (x-y)\vc S(x)\cdot\vc S(y)
\ee
describes (effective) exchange interactions between the spins of electrons. In (\ref{3.8}), $\hat t(x-y)$ is the amplitude for hopping of an electron from site $y$ to site $x$. Selfadjointness of the Hamiltonian $H$ implies that $\hat t(x-y)=\overline{\hat t(y-x)}$.
The constant $U_0$ is a measure of the strength of the on-site repulsion, and $J_{\neq}(x-y)$ is the exchange coupling between spins at $x$ and $y$, with $J_{\neq}(0)=0$.

The term $U$ can be rewritten as follows. When $x$ is empty or singly occupied $U_0n_+(x)n_-(x)$ vanishes; when $x$ is doubly occupied it takes the value $U_0$. Thus, by (\ref{3.6}),
\bb{normal}
 U_0n_+(x)n_-(x)=\frac{1}{2}U_0n(x)-\frac{2}{3}U_0\vc S(x)\cdot \vc S(x)=\frac{2}{3}U_0:\vc S(x)\cdot \vc S(x):\,,
\ee where $:\, .\, :$ denotes normal ordering. We now set
$J(0)=\frac{4}{3}U_0$, $J(x)=J_{\neq}(x)$, for $x{\neq} 0$, and
find \bb{3.11} H=T+E\,, \ee with \bb{3.12}
E=\frac{1}{2}\Sum{x,y}J(x-y):\vc S(x)\cdot\vc S(y):\,. \ee

The {\em Heisenberg model} corresponds to a completely flat band, i.e.,
\bb{3.13}
T=0\,,\quad \text{or}\quad \hat t(x)\equiv 0\,.
\ee

Our goal is to study the grand-canonical partition function
\bb{3.14}
\Xi (\beta,\mu):=\tr \left(e^{-\beta[H-\mu\Sum xn(x)]}\right)
\ee
and the thermal equilibrium state given by the density matrix
\bb{3.15}
P_{\beta,\mu}=\Xi(\beta,\mu)^{-1}e^{-\beta[H-\mu \Sum xn(x)]}
\ee
for the family of Hamiltonians introduced above.

{\em Remark.} Of course, these objects must first be calculated for arbitrary bounded regions, $\Lambda$, in the lattice $\Gamma$ (indicated by a superscript ``$\Lambda$'').
Afterwards, one may attempt to pass to the thermodynamic limit, $\Lambda \nearrow \Gamma$, setting
\bb{3.16}
\beta g(\beta,\mu)=-\lim_{\Lambda \nearrow \Gamma} \frac{1}{|\Lambda|}\ln \Xi^\Lambda(\beta,\mu)
\ee
and
\bb{3.17}
\omega_{\beta,\mu}(\cdot)=\lim_{\Lambda \nearrow \Gamma}\tr\left(P_{\beta,\mu}^\Lambda(\cdot)\right)\,,
\ee
where $|\Lambda|$ is the number of sites in $\Lambda$. These are standard matters (see e.g. \cite{10}) and will not be discussed any further. The superscript ``$\Lambda$'' will often be suppressed in our notation.

It is convenient to study $\Xi(\beta,\mu)$ and the state $\omega_{\beta,\mu}$ by making use of {\em functional integration}. With every $x\in\Gamma$ and every imaginary time $t\in[0,\beta)$, we associate anticommuting (Grassmann) variables, $\bar C_s(x,t)$ and $C_s(x,t)$ with
$$
\{ \overset{(-)}{ C_s}(x,t)\,,\,\overset{(-)}{ C_{s'}}(x',t')\}=0\,,
$$
and \bb{3.18} C_s^{\#}(x,t+\beta)=-C_s^{\#}(x,t)\,. \ee The
anti-periodic boundary conditions in (\ref{3.18}) are a
consequence of the KMS condition; see e.g. \cite{11}. We introduce
an (imaginary-time) action functional
\bb{3.19}
S_J (\bar{C},C) = \int_0^{\beta} d t \, \left[ \left( \sum_{x,s} \bar{C}_{s}
  (x,t) \frac{\partial}{\partial t} C_s (x,t) \right) - H(\bar{C}(.,t), C(.,t))
-\mu \sum_{x,s} \bar{C}_s (x,t) C_{s} (x,t) \right] ,
\ee
where $H (\bar{C},C)$ is obtained from the Hamiltonian $H$ by replacing the operators $c_s (x)$ and $c^\dag_s (x)$ by $C_s (x,t)$
and $\bar{C}_s (x,t)$, respectively. Because of the presence of products of $C$- and $\bar C$-fields at coinciding sites and imaginary times the quartic term needs to be regularized; (the ambiguity of the quadratic terms is an unimportant constant). We introduce an infinitesimal separation of the imaginary times, replacing $\vc{S}(x)$ in (\ref{3.12}) by
\bb{3.24reg}
\vc S^{\epsilon}(x,t)=\frac{1}{2}\Sum{s,s'}\bar C_s(x,t+\epsilon){\bf \sigma}_{ss'}C_{s'}(x,t-\epsilon)\,
\ee
in the action. Then the time ordering of the $C$- and $\bar C$-fields corresponds to the normal ordering of the operators
$c^{\sharp}$. Because of (\ref{3.12}) the regularized action is now given by
\begin{multline}
S_J^{\epsilon}(\bar C,C)=\int\limits_0^\beta dt\bigg[\left(\Sum{x,s}\bar
C_s(x,t)\frac{\partial}{\partial t}C_s(x,t)\right)
-\Sum{x{\neq} y}\hat t(x-y)\Sum{s}\bar C_s(x,t)C_s(y,t) \\
+\frac{1}{2}\Sum{x,y}J (x-y)\vc S^{\epsilon}(x,t)\cdot\vc S^{\epsilon} (y,t)
+\mu\Sum{x,s}\bar C_s(x,t)C_s(x,t) \bigg]\,,
\end{multline}

Berezin's integration form for anticommuting variables is formally
given by \bb{3.20} {\mathcal D}\bar C{\mathcal
D}C=\Pi_{t\in[0,\beta)}\Pi_x\Pi_s d\bar C_s(x,t)dC_s(x,t)\,. \ee
Then \bb{3.21} \Xi(\beta,\mu)=\lim_{\epsilon \searrow
0}\int\mathcal D\bar C\mathcal DC e^{S^{\epsilon}_J(\bar C,C)} \ee
and the state $\omega_{\beta, \mu}$ can be reconstructed from the
imaginary-time Green functions \bb{3.22} \langle\Pi_j \bar C_{s_j}
(x_j,t_j)
C_{s'_j}(x'_j,t'_j)\rangle_{\beta,\mu}=\Xi(\beta,\mu)^{-1}
\lim_{\epsilon \searrow 0}  \int\mathcal D\bar C\mathcal DC
e^{S^{\epsilon}_J(\bar C,C)}\Pi_j\bar C_{s_j}(x_j,t_j)
C_{s'_j}(x'_j,t'_j)\,; \ee see, e.g., \cite{11}, \cite{12}.

If $S^{\epsilon}_J(\bar C,C)$ were quadratic in $C$ and $\bar C$
the Berezin integrals in (\ref{3.21}), (\ref{3.22}) could be
evaluated and expressed in terms of determinants. The only
contribution to $S_J^{\epsilon}(\bar C,C)$ {\em not} quadratic in
$\bar C$, $C$ comes from the term \bb{3.23} \int\limits_0^\beta dt
\Sum{x,y}J(x-y)\vc S^{\epsilon}(x,t)\cdot \vc S^{\epsilon}(y,t)\,
. \ee

By introducing a Lagrange-multiplier field (``Hubbard-Stratonovich
transformation'') this term, too, can be rendered quadratic in
$\bar C$ and $C$. The Lagrange multiplier field will turn out to
be the {\it Weiss exchange field} $\vc W$.

Let $K$ denote the matrix inverse of $J$, i.e., $\hat K(k)=\hat
J(k)^{-1}$, where $k$ is a point in the first Brillouin zone,
$B_\Gamma$, of the lattice $\Gamma$, and $\hat J$, $\hat K$ denote
the Fourier transforms of $J$ and $K$. If $U_0$ is chosen to be
large enough then $J$ and $K$ are positive-definite, i.e., $\hat
J(k)>0$, $\hat K(k)>0$, for $k\in B_\Gamma$. We introduce a real,
vector-valued random field, $\vc W(x,t)$, $x\in\Gamma$,
$t\in[0,\beta)$, with \bb{3.25} \vc W(x,t+\beta)=\vc W(x,t) \ee
(KMS condition for bosons; see e.g. \cite{11}), assumed to have a
Gaussian distribution \bb{3.26} d\mu_K(\vc W)=e^{S_K(\vc
W)}\mathcal D \vc W\,, \ee where \bb{3.27} S_K(\vc
W)=-\frac{1}{2}\int\limits_0^\beta dt\Sum{x,y}\vc W(x,t)K(x-y)\vc
W(y,t)\,, \ee \bb{3.28} \mathcal D\vc
W=\const_K\Pi_{t\in[0,\beta)}\Pi_x d^3W(x,t)\,, \ee and $\const_K$
is chosen such that \bb{3.29} \int d\mu_K(\vc W)=1\,. \ee We
define\footnote{Henceforth we suppress the regularization indicated by $\epsilon$.}
\bb{3.30} S_0(\bar C,C)=S_{J=0}(\bar C,C)\,, \ee \bb{3.31}
S_1(\bar C,C;\vc W)=-\int\limits_0^\beta dt\left(\Sum x \vc
W(x,t)\cdot\vc S(x,t)\right)\,, \ee and \bb{3.32} \bar S(\bar
C,C;\vc W)=S_0(\bar C, C)+S_1(\bar C, C;\vc W)+S_K(\vc W)\,. \ee
We note that $\bar S(\bar C,C;\vc W)$ is quadratic in $\bar C$ and
$C$ and in $\vc W$ (separately), and that \bb{3.33} \int \mathcal
D \vc W e^{\bar S(\bar C, C; \vc W)}=e^{S_J(\bar C,C)}\,, \ee as
follows by quadratic completion.

It is apparent from (\ref{3.31}) that $\vc W(x,t)$ plays
the role of the Weiss exchange field. In order to calculate the
grand canonical partition function $\Xi(\beta,\mu)$, we have to
integrate the R.S. of (\ref{3.33}) over $\bar C$ and $C$. Since
$\bar S(\bar C,C;\vc W)$ is quadratic in $\bar C$ and $C$, it is
tempting to interchange the $\vc W$ - and the $\bar C$, $C$ -
integrations. First doing the $\bar C$, $C$ - integration yields
\bb{3.34} \int\mathcal D\bar C \mathcal D C e^{S_0(\bar
C,C)+S_1(\bar C,C;\vc W)} = \det (D_\vc W)\,, \ee where the
operator $D_\vc W$ acts on the space \bb{3.35} {\mathfrak
h}=C^\infty([0,\beta))_{ap}\otimes \left(l^2(\Gamma)\otimes
\mathbb C^2\right) \ee (``ap'' stands for ``antiperiodic'') and is
determined by requiring that
$$
S_0(\bar C,C)+S_1(\bar C,C;\vc W)=\int\limits_0^\beta dt \sum\limits_{x,s}\bar C(x,t)(D_\vc W C)(x,t)\,.
$$
Thus, $D_\vc W$ is given by
\bb{3.36}
D_\vc W=\frac{\partial}{\partial t}\otimes \identy +\identy\otimes \left(\hat t -\mu\identy\right) +\frac{1}{2} \vc W(x,t)\cdot \pmb{\sigma}\,,
\ee
where $(\hat t h)(x)=\sum\limits_y \hat t(x-y)h(y)\,,\quad h\in l^2(\Gamma)$.

After Fourier transformation in $t$ and $x$, vectors in the space $\mathfrak h$ are given by two-component spinors
$$
\phi(\vc k,k_0)=\begin{pmatrix} f_+(\vc k,k_0) \\ f_-(\vc k,k_0) \end{pmatrix}\,,
$$
with $\vc k\in B_\Gamma$ and $k_0 = (\pi/\beta)(2n+1)$, $n\in
\mathbb Z$, and \bb{3.37} (D_\vc W \phi)(\vc k,k_0)=\left(-ik_0+
t(\vc k)-\mu\right)\phi(\vc
k,k_0)+\frac{1}{2}\int\limits_{B_\Gamma} \frac{d\vc
k'}{(2\pi)^3}\sum\limits_{k'_0}\hat{\vc W}(\vc k-\vc
k',k_0-k'_0)\cdot(\pmb{\sigma}\phi)(\vc k',k'_0)\,, \ee where
$t(\vc k)$ denotes the Fourier transform of $\hat t(x)$. Defining
the {\em effective action}, $S_{eff}(\vc W)$, of the exchange
field, $\vc W$, by \bb{3.38} S_{eff}(\vc W)=\ln\det(D_\vc W) +
S_K(\vc W)\,, \ee we find that, for example, \bb{3.39}
\Xi(\beta,\mu)=\int \mathcal D\vc W e^{S_{eff}(\vc W)}\,. \ee Note
that \bb{3.40} \frac{\delta}{\delta\vc W(x,t)}e^{\bar S(\bar
C,C;\vc W)}=-\left(\sum\limits_y K(x-y)\vc W(y,t)+\vc
S(x,t)\right) e^{\bar S(\bar C,C;\vc W)}. \ee Setting
$$
\delta_{\vc W(x,t)}=\frac{\delta}{\delta \vc W(x,t)}+\sum\limits_y K(x-y)\vc W(y,t)\,,
$$
the imaginary-time Green function corresponding to a product, $\prod\limits_{j=1}^n S_{\alpha_j}(x_j,t_j)$, of spin operators turns out to be given by
\bb{3.41}
\langle \prod\limits_{j=1}^n S_{\alpha_j}(x_j,t_j)\rangle_{\beta,\mu}=(-1)^n\Xi(\beta,\mu)^{-1}\int \mathcal D \vc W \prod\limits_{j=1}^n \delta_{W_{\alpha_j}(x_j,t_j)}e^{S_{eff}(\vc W)}\,,
\ee
where we have used (\ref{3.38}) and (\ref{3.40}).

Our purpose is now to determine $S_{eff}(\vc W)$ as explicitly as possible, in order to get some insight into the R.S. of (\ref{3.39}) and (\ref{3.41}).
 \label{chap3}
\section{Effective $\sigma$-models of the exchange field}
\label{sec:4}\setcounter{equation}{0}

If we want to apply the method of steepest descent to estimate the R.S. of (\ref{3.39}) and of (\ref{3.41}), we must look for the (absolute and local) maxima of $S_{eff}(\vc W)$. This leads us to consider mean-field ans\"atze for configurations, $\vc W$, corresponding to local maxima of $S_{eff}(\vc W)$.

\subsection*{Ferromagnetic mean-field theory}
We set
\bb{3.42}
\vc W(x,t)=W_0\vc n\,,
\ee
independently of $x$ and $t$, where $\vc n$ is a unit vector, and $W_0>0$.
Because of the symmetry of the problem under rotations of $\vc W$, we can choose $\vc n=\vc n_z$ to point into the $z$-direction.
The problem of evaluating
\bb{3.43}
g^{mf}(\beta,\mu;W_0)=\lim_{\Lambda\nearrow\Gamma}-\frac{1}{\beta|\Lambda|}S_{eff}^\Lambda (\vc W\equiv W_0\vc n_z)
\ee
is elementary, because, in momentum space, all modes decouple from one another, and we can appeal to well known calculations (see e.g. \cite{13}, \cite{11}) to obtain the answer:
\begin{multline}\label{3.44}
g^{mf}(\beta,\mu;W_0)=\frac{1}{2}\hat K(0)W_0^2-\lim_{\Lambda\nearrow\Gamma}\frac{1}{\beta|\Lambda|}\ln\det(D_{\vc W=W_0\vc n})\\
=\frac{1}{2}\hat K(0)W_0^2 -\frac{1}{\beta}\int\limits_{B_\Gamma}\frac{d\vc k}{(2\pi)^3}\ln\Big[\left(1+e^{-\beta(\epsilon(\vc k)+W_0/2)}\right)\times\\
\times\left(1+e^{-\beta(\epsilon(\vc k)-W_0/2)}\right)\Big]+\const.\,,
\end{multline}
where
\bb{3.45}
\epsilon(\vc k)= t(\vc k)-\mu\,.
\ee
Formula (\ref{3.44}) can also be derived directly from the definition of $\ln\det(D_{\vc W})=\trs\ln(D_\vc W)$ by using the Poisson summation formula.

\begin{figure}[h]
\begin{center}
\input{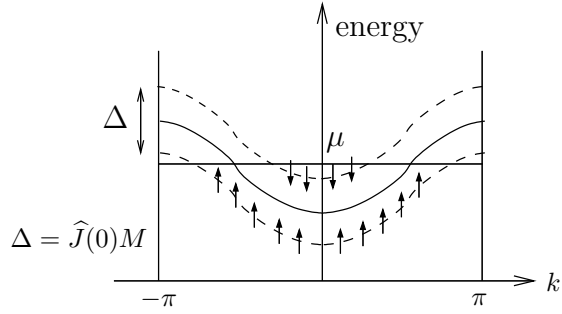}
\end{center}
\caption{Spin bands}\label{figure3}
\end{figure}

In the limit $\beta\longrightarrow\infty$, expression (\ref{3.44}) converges to
\begin{equation}\label{3.46}
\begin{split}
g^{mf}(\infty,\mu;W_0) = \, &\frac{1}{2}\hat K(0)W_0^2 \\
&+\int\limits_{B_\Gamma}\frac{d\vc k}{(2\pi)^3}\left[2\epsilon(\vc k)\Theta\left(-\frac{W_0}{2}-\epsilon(\vc k)\right)+\left(\epsilon(\vc k)-\frac{W_0}{2}\right)\Theta\left(\frac{W_0}{2}-|\epsilon(\vc k)|\right)\right]\\
&+\const.\,,
\end{split}
\end{equation}
which is the ground state energy density of electrons in a constant exchange field $\vc W=W_0\vc n_z$. The value of $W_0$ minimizing $g^{mf}$ is found by solving the equation $(\partial/\partial W_0)g^{mf}(\beta,\mu;W_0)=0$, i.e.,
\bb{3.47}
\hat K(0)W_0=\frac{1}{2}\int\limits_{B_\Gamma}\frac{d\vc k}{(2\pi)^3}\left[(e^{\beta(\epsilon(\vc k)-\frac{W_0}{2})}+1)^{-1}-(e^{\beta(\epsilon(\vc k)+\frac{W_0}{2})}+1)^{-1}\right]\,.
\ee
When $\beta\rightarrow 0$ the R.S. of (\ref{3.47}) tends to $0$, and we conclude that, for small $\beta$, this equation only has the trivial solution, $W_0=0$. When $\beta\rightarrow \infty$, equation (\ref{3.47}) yields the equation
\bb{3.48}
\hat K(0)W_0=\frac{1}{2}\int\limits_{B_\Gamma}\frac{d\vc k}{(2\pi)^3}\Theta\left(\frac{W_0}{2}-|\epsilon(\vc k)|\right)\,.
\ee
Besides the trivial solution, this equation also has a non-trivial solution $W_0>0$, provided $\hat K(0)$ is sufficiently small (depending on $\hat t$). We recall that
\bb{3.49}
\hat K(0)=\hat J(0)^{-1}=\left(\frac{4U_0}{3}+\hat J_{\neq}(0)\right)^{-1}\,.
\ee
Thus, for $U_0/t_*$, with $t_*:=\max|\hat t(x)|$, large enough, equation (\ref{3.47}) has a non-trivial solution, provided $\beta$ is large enough, {\em even} in the Hubbard model, where $J_{\neq}=0$. Note that, by (\ref{3.40}) and (\ref{3.41}),
\bb{3.50}
-\Sum{y} K(x-y)\langle\vc W(y,t)\rangle_{\beta,\mu}=\langle\vc S(x,t)\rangle_{\beta,\mu}\,.
\ee
Thus, in mean-field theory,
\bb{3.51}
\hat K(0)W_0=M\,,
\ee
where $W_0$ is the non-trivial solution of equation (\ref{3.47}), and $M$ is the spontaneous magnetization in mean-field theory (i.e., $-M\vc n_z=\langle\vc S(x,t)\rangle^{mf}_{\beta,\mu}$). Thus, equation (\ref{3.48}) tells us that, at zero temperature ($\beta\rightarrow\infty$),
\bb{3.52}
M=\frac{1}{2}\int\limits_{B_\Gamma}\frac{d\vc k}{(2\pi)^3}\Theta\left(\frac{\hat J(0)}{2}M-|\epsilon(\vc k)|\right)\,,
\ee
where $\hat J(0)=\Sum{x}J(x)=\hat K(0)^{-1}$. This equation has an obvious interpretation apparent from Fig. \ref{figure3}.

Matters simplify further for a {\em flat band}, $t\equiv 0$, i.e., for the Heisenberg model. Then eq. (\ref{3.48}) reduces to
\bb{3.53}
M=\left\lbrace\begin{array}{l}
        1/2\,,\quad -\hat J(0)/4<\mu<\hat J(0)/4\,,\\
        0\,,\quad \text{otherwise}\,,
\end{array}
\right.
\ee
for $\beta\rightarrow\infty$.
For, eq. (\ref{3.46}) implies that
\bb{3.54}
g^{mf}(\infty,\mu;W_0)=\frac{1}{2}\hat K(0)W_0^2-\frac{W_0}{2}\Theta\left(\frac{W_0}{2}-|\mu|\right)\,,
\ee
up to an unimportant constant. Thus, for $|\mu|<1/(4\hat K(0))=\hat J(0)/4$, $g^{mf}(\infty,\mu;W_0)$ has a quadratic minimum at $W_0=1/(2\hat K(0))$, with
$$
g^{mf}\left(\infty,\mu;W_0=\frac{1}{2\hat K(0)}\right)=-\frac{1}{2\hat K(0)}\,,
$$
and
\bb{3.55}
\frac{\partial^2g^{mf}}{\partial W_0^2}\left(\infty, \mu;W_0\approx \frac{1}{2\hat K(0)}\right)\equiv \hat K(0)\,.
\ee

\subsection*{Antiferromagnetic mean-field theory}
Antiferromagnetism and N\'eel ordering in Hubbard models is
discussed in some detail in \cite{13}, so we shall be
brief. To simplify matters, we consider the lattices $\Gamma
= \mathbb Z^d$, $d=1,2,3,\dots$. Our mean-field ansatz
configuration for the exchange field $\vc W$ is given by
\bb{3.62}
\vc W(x,t)=(-1)^{|x|}W_0\vc n_z\,,
\ee
which describes N\'eel order. We
skip detailed calculations and give the result for the free energy
\begin{multline}\label{3.66}
g^{mf}(\beta,\mu,W_0)=\frac{1}{2}\hat K(\vc
k_\pi)W_0^2-\frac{1}{\beta}\int\limits_{B_\Gamma}\frac{d\vc
k}{(2\pi)^3}\ln\left[\left(1+e^{-\beta\sqrt{\epsilon(\vc
k)^2+(1/4)W_0^2}}\right)\left(1+e^{\beta\sqrt{\epsilon(\vc
k)^2+(1/4)W_0^2} }\right)\right]\\
+\const\,,
\end{multline}
where $\vc k_\pi=(\pi,\pi,\dots,\pi)$ denotes a corner of the Brillouin zone.
In order to derive (\ref{3.66}), we assume that $\epsilon(\vc
k+\vc k_\pi)=-\epsilon(\vc k)$, i.e., we assume the band to be
{\em half-filled}. Note that small deviations $\delta \epsilon(\vc
k)=\epsilon(\vc k+\vc k_\pi) + \epsilon(\vc k)$ from half filling
can be taken into account with the help of perturbation theory in
$\delta\epsilon(\vc k)$. The idea behind the derivation of (\ref{3.66})
is to consider simultaneously the contributions of $\vc{k}$ and
$\vc k + \vc{k}_{\pi}$, diagonalising a $4\times 4$ matrix.
The energy eigenvalues, for a fixed $\vc k$, are given by $\epsilon_{\pm} (\vc k) =
\pm \sqrt{\epsilon^2 (\vc k) + (1/4) W_0^2}$, each one with twofold degeneracy.
{F}rom (\ref{3.66}) we obtain the
mean-field value of $W_0$ by solving the equation
$\partial g^{mf}(\beta,\mu; W_0)/\partial W_0=0$. In the limit
where $\beta\rightarrow\infty$, this equation reduces to
\bb{3.67}
\hat K(\vc k_\pi)=\frac{1}{4}\int\limits_{B_\Gamma}\frac{d\vc
k}{(2\pi)^3}\frac{1}{\sqrt{\epsilon(\vc k)^2+(1/4)W_0^2}}\,.
\ee
The R.S.
diverges logarithmically when $W_0\rightarrow 0$, for a lattice $\Gamma$ of arbitrary dimension. Thus
(\ref{3.67}) has a non-trivial solution $W_0>0$ for arbitrarily
large values of $\hat K(\vc k_\pi)=(4U_0/3 +J_{\neq}(\vc
k_\pi))^{-1}$, i.e., for arbitrarily small values of the on-site
repulsion $U_0$ (even if $J_{\neq}=0$), and hence the same is true for finite, but
sufficiently large values of $\beta$. Thus
\bb{3.68}
|\epsilon_{\pm}(\vc k)|\geq \frac{1}{2}W_0>0\,,
\ee
and we conclude that a strictly positive {\em energy gap $W_0>0$ opens at the
Fermi-surface}.

We note that the solution of (\ref{3.67}) behaves like
\bb{3.69}
W_0\varpropto t_*e^{-\alpha\hat K(\vc k_\pi)t_*}\,,
\ee
for some constant $\alpha$. For the half-filled Hubbard model ($J_{\neq}=0$, $U_0>0$), $\hat K(\vc k_\pi)=3/(4U_0)$, and we find that $W_0$ has an essential singularity at $U_0=0$.

Comparing these results with those obtained for the ferromagnetic mean-field ansatz, we conclude that, at zero temperature, in the Hubbard model with $\hat K(0)=\hat K(\vc k_\pi)=3/(4U_0)$, a non-trivial ferromagnetic mean-field solution does {\em not} exist when $U_0$ is small, while an antiferromagnetic mean-field solution exists {\em for arbitrarily small values} of $U_0$ at {\em half filling}, where the (mean-field) ground state energy of the N\'eel state is found to be {\em below} the ground state energy of the ferromagnetic state.

\subsection*{The magnetic $\sigma$-model}
As we have seen, these ans\"atze correspond to quadratic (local) maxima of the effective action as a function of the {\em length $W_0$ of the exchange field}. This suggests that the effective action $S_{eff}(\vc W)$ is maximal on exchange-field configurations of constant length, $|\vc W(x,t)|\approx W_0$. It may therefore be appropriate to study the effective action (\ref{3.38}) with the constraint
\bb{4.51}
 |\vc W(x,t)| = W_0\,,\quad \text{for all}\quad t\in [0,\beta)\,,\quad x\in\Gamma\,.
\ee
To study magnetic order, we consider the following {\em ansatz} for the effective action on configurations of exchange fields of constant length, $|{\vc W}(x,t)|=W_0$:
\begin{equation}\label{Fsm}
\begin{split}
\tilde S_{eff}(\vc W)= &\int\limits_0^\beta dt\Sum{x}\Big\{-\frac{W_0^2}{2}\Sum{y}\hat{\vc W}(x,t)K(x-y)\hat{\vc W}(y,t)+C_t|\partial_t \hat{\vc W}(x,t)|^2+C_{\nabla}|\nabla \hat{\vc W}(x,t)|^2\Big\}\\
&+C_{WZ}\Int{S^{2,N}}dt ds\Sum{x}\hat{\vc W}(x,t,s)\cdot(\partial_{t}\hat{\vc W}(x,t,s)\wedge \partial_{s}\hat{\vc W}(x,t,s))+\text{higher order terms}\,,
\end{split}
\end{equation}
where we have set $\hat{\vc W}(x,t):=W_0^{-1}{\vc W}(x,t)$ and where, for each $x\in\Gamma$, $\hat{\vc W}(x,t,s)$ is chosen to be a smooth extension of $\hat{\vc W}(x,t)$ from the circle $\{t|0\leq t\leq\beta\}$ to the northern hemisphere, $S^{2,N}$, of the $2$-sphere of radius $\beta/(2\pi)$.
The operator $\nabla$ denotes the finite difference gradient on the lattice, and the corresponding term in the effective action has to be interpreted as
$$
\Sum{x}|\nabla\hat{\vc W}(x,t)|^2:=\Sum{|x-y|=1}|\hat{\vc W}(y,t)-\hat{\vc W}(x,t)|^2\,.
$$

The functional $\tilde S_{eff}({\vc W})$ is the (imaginary-time) action of a {\em non-linear $\sigma$-model} with $SO(3)$ symmetry and with a Wess-Zumino term. This model is called {\em magnetic $\sigma$-model}. It is expected to describe exchange field configurations $\vc W$ in thermal equilibrium at inverse temperature $\beta$, provided the coefficients $C_t$, $C_\nabla$ and $C_{WZ}$ are chosen in such a way that
\bb{4.95'}
\tilde S_{eff}({\vc W})\approx S_{eff}({\vc W})\arrowvert_{|W(x,t)|=W_0}+\operatorname{const}\,.\,,
\ee
on configurations $\vc W(x,t)$ satisfying the constraint $|\vc W(x,t)|=W_0$, with $S_{eff}(\vc W)$ as in (\ref{3.38}).
More precisely, if ${\vc W}={\vc W}_{mf}+{\pmb \chi}$, where ${\vc W}_{mf}$ is a mean(-exchange)-field configuration and ${\pmb\chi}$ is a small perturbation of ${\vc W}_{mf}$, i.e., ${\pmb\chi}(x,t)\perp{\vc W}_{mf}(x)$ and $|{\pmb\chi}(x,t)|\ll 1$, for all $x\in\Gamma$ and all times $t$, then we want (\ref{4.95'}) to hold {\em exactly} to {\em second order} in ${\pmb\chi}$. Eq. (\ref{Fsm}) makes it clear that, in order to calculate $C_t$, $C_\nabla$ and $C_{WZ}$ such that (\ref{4.95'}) holds to second order in ${\pmb\chi}$, it suffices to consider {\em time-independent} exchange fields, ${\vc W}(x,t)\equiv{\vc W}(x)$, to calculate $C_\nabla$, while it suffices to consider {\em site-independent} exchange fields, ${\vc W}(x,t)\equiv{\vc W}(t)$, when calculating $C_t$ and $C_{WZ}$.
Obviously, the {\em signs} of $C_t$ and $C_\nabla$ will determine the {\em stable} mean-field configuration ${\vc W}_{mf}$. For ${\vc W}_{mf}$ to be {\em independent} of time $t$, $C_t$ must be {\em negative} (with our sign convention).
For {\em ferromagnetism} to prevail, $C_\nabla$ must be {\em negative}.

In the next two subsections, the coefficients $C_t$ and $C_\nabla$ are determined  in such a way that (\ref{4.95'}) holds to second order in ${\pmb\chi}$, for an {\em appropriate choice of} ${\vc W}_{mf}$ (depending on our choice of parameters in the original model). The coefficient $C_{WZ}$ of the Wess-Zumino term will be calculated and shown to be {\em imaginary} in the next to last subsection of Section 4. The Wess-Zumino term is crucial in arriving at the correct {\em magnon dynamics} for {\em ferromagnetically ordered} magnets.

\subsection*{Stability of antiferromagnetic ordering}
In this subsection we calculate the coefficients $C_t$ and $C_{\nabla}$ of the
magnetic $\sigma$-model (\ref{Fsm}), in such a way that (\ref{4.95'}) holds, in a regime where these coefficients can be calculated perturbatively. For the {\em Hubbard model} this is the regime where the band is {\em half-filled} and the signs of the coefficients $C_t$ and $C_{\nabla}$ will indeed determine the {\em stable} mean exchange field to be time-independent and {\em antiferromagnetically} ordered.

We first derive an innocent looking, but important identity, which will also be useful in the next section.
We set
\bb{4.6}
D_{\vc W}^-=-\partial_t+\epsilon -W\,.
\ee
Equation (\ref{3.36}) shows that
$$
D_{\vc W}^T=-\partial_t+\epsilon+\frac{1}{2}(W_1\sigma_1-W_2\sigma_2+W_3\sigma_3)\,,
$$
where the superscript ``$T$'' stands for transposition, (i.e., taking the adjoint, followed by complex conjugation). Hence
\bb{4.7}
\sigma_2D_{\vc W}^T\sigma_2=-\partial_t+\epsilon-W=D_{\vc W}^-\,.
\ee
Furthermore,
\bb{4.8}
D_{\vc W}^*=-\partial_t+\epsilon+W=D_{-\vc W}^-\,.
\ee
Since $\sigma_2=\sigma_2^*=\sigma_2^{-1}$, it follows from (\ref{4.7}) that
\bb{4.9}
\det(D_{\vc W})=\det(D_{\vc W}^T)=\det(\sigma_2 D_{\vc W}^T\sigma_2)=\det(D_{\vc W}^-)\,,
\ee
and hence
\bb{4.10}
\overline{\det(D_{\vc W})}=\det(D_{\vc W}^*)=\det(D_{-\vc W}^-)=\det(D_{-\vc W})\,.
\ee
This identity shows that the real part of the logarithm of the determinant of $D_{\vc W}$ is given by:
\begin{align*}
\Re\ln\det(D_{\vc W})&=\frac{1}{2}\ln\det(D_{\vc W}D_{-\vc W})\,,\\
D_{\vc W}D_{-\vc W}&=(\partial_t+\epsilon)^2-\frac{W_0^2}{4}-[\epsilon,W]-(\partial_t{W})\,.
\end{align*}
We are tempted to treat $[\epsilon,W]+(\partial_t{W})$ as a perturbation of 
\bb{4.109'}
\Lambda:=(\partial_t+\epsilon)^2-\frac{W_0^2}{4}\,.
\ee
This is justified if $|\epsilon|\ll W_0$ and $|\partial_t W|\ll W_0$. In this regime the
band of electrons with spin parallel to the background field, ${\vc W}_{mf}$,
is filled and the band of electrons with spin anti-parallel to ${\vc W}_{mf}$ is empty, i.e., the lattice is
half-filled. At half-filling, we expect the {\em Hubbard model} to exhibit an
antiferromagnetic phase at sufficiently low temperatures.
We recall that $\ln\det=\trs\ln$ and expand the logarithm up to quadratic order in $[\epsilon,W]/\Lambda$ and $(\partial_t W)/\Lambda$. Terms linear in $W$ vanish at critical points of the effective action; $\vc W$-independent terms are omitted. We are then left with
\bb{Alndet1}
\Re\ln\det(D_{\vc W})=-\frac{1}{4}\trs(\Lambda^{-1}([\epsilon,W]+(\partial_t W))\Lambda^{-1}([\epsilon,W]+(\partial_t W)))+\text{h.o.}\,,
\ee
where $\text{h.o.}$ stands for terms of higher order in $|\epsilon|/W_0$ or $|\partial_t W|/W_0$.
Next, we pull $\Lambda^{-1}$ through $[\epsilon,W]+\partial_tW$ using the identity
\bb{pullthrough}
[\Lambda^{-1},[\epsilon,W]+(\partial_tW)]=-\Lambda^{-1}[\Lambda,[\epsilon,W]+(\partial_tW)]\Lambda^{-1}\,,
\ee
which shows that the commutator only
contributes higher-order terms to (\ref{Alndet1}). Mixed terms with one spatial and one time derivative of $W$ do not contribute to the trace. We thus arrive at
\bb{Alndet}
\Re\ln\det(D_{\vc W})=-\frac{1}{4}\trs(\Lambda^{-2}([\epsilon,W]^2+(\partial_t W)^2))+\text{h.o.}\,.
\ee
Comparison with (\ref{Fsm}) then yields an expression for $C_{\nabla}$,
$$
C_{\nabla}=\frac{W_0^2}{8(2\pi)^{3}}\Int{B_{\Gamma}}d\vc k|\nabla\epsilon(\vc k)|^2\frac{1}{\beta}\Sum{k_0}\frac{1}{((-ik_0+\epsilon(\vc k))^2-W_0^2/4)^2}\overset{\beta W_0\rightarrow\infty}{\longrightarrow}\frac{1}{4(2\pi)^{3}W_0}\Int{B_{\Gamma}}d\vc k|\nabla\epsilon(\vc k)|^2\,,
$$
which is positive. Thus the real part of the $\sigma$-model action (\ref{Fsm}) for the Hubbard-model exhibits its maximum (w.r.t. time-independent exchange fields) on configurations for which
$$
\int\limits_0^\beta dt\Sum{x}|\nabla\hat{\vc W}(x,t)|^2
$$
is {\em maximal}, and this is the case for a N\'eel ordered (staggered) exchange field.

{F}rom (\ref{Alndet}) we can also read off the $C_t$-coefficient in (\ref{Fsm}) to be given by
$$
C_t=-\frac{W_0^2}{8}\frac{1}{\beta}\Sum{k_0}\frac{1}{((-ik_0+\epsilon(\vc k))^2-W_0^2/4)^2}\overset{\beta W_0\rightarrow\infty}{\longrightarrow}-\frac{1}{4W_0}\,,
$$
which is manifestly negative. 
From these calculations and the mean field equation (\ref{3.67}) we conclude that the {\em Hubbard model at half filling} exhibits a {\em stable} antiferromagnetic phase if the on-site repulsion $U_0$ is large enough as compared to the hopping term, and if the temperature is sufficiently low.

\subsection*{Linear stability of the ferromagnetic mean exchange field}
In this subsection we analyze conditions that imply stability of {\em ferromagnetically} ordered exchange fields. In the {\em Hubbard model}, we expect a stable ferromagnetic phase to prevail when the band is weakly filled and fairly flat. In such a regime, since there is no energy gap at the Fermi surface, the operator $\Lambda$ defined in (\ref{4.109'}) has zeros and the perturbative expansion of the last subsection breaks down. We therefore have to resort to a linear stability analysis around a ferromagnetically ordered mean exchange field. We will see that the stability of the ferromagnetically ordered exchange field is due to contributions close to the Fermi surface.

The stability of a ferromagnetically ordered exchange field w.r.t. $x$-independent fluctuations can be shown quite easily, since the operator $D_{\vc W}$ is diagonal in momentum space when $\vc W(x,t)=\vc W(t)$ is $x$-independent. We can then write
\bb{decomp}
\lim_{\Lambda \nearrow\Gamma}\frac{1}{|\Lambda|}\ln\det(D_{\vc W})=\Int{B_{\Gamma}}\frac{d\vc p}{(2\pi)^3}\ln\det\left(\partial_t +\hat\epsilon(\vc p)+\frac{\vc W(t)}{2}\cdot\pmb{\sigma}\right)\,.
\ee
In the operator formalism outlined at the beginning of Section \ref{chap3},
$$
\det\left(\partial_t + \hat\epsilon(\vc p) + \frac{\vc W(t)}{2}\cdot \pmb{\sigma}\right)=\const\times\tr_{{\mathbb C}^4}U_{\epsilon,\vc W}(\beta ,0)\,,
$$
where
\begin{align}
\partial_t U_{\epsilon,\vc W}(t,s)&=h_{\epsilon,\vc W}\,(t)\,U_{\epsilon,\vc W}(t,s)\,,\\
U_{\epsilon,\vc W}(t,t)&=\identy_4\,,
\end{align}
and
$$
h_{\epsilon,\vc W}\,(t)\,=\begin{pmatrix}
0 & 0 \quad\quad\quad 0 & 0 \\
\begin{array}{c}0\\0\end{array} & \hat \epsilon(\vc p)\identy_2+\frac{\vc W(t)}{2}\cdot\pmb{\sigma} & \begin{array}{c}0\\0\end{array}\\
0 & 0 \quad\quad\quad 0& 2\hat\epsilon(\vc p)
\end{pmatrix}\,.
$$
The trace of the propagator $U_{\epsilon,\vc W}(\beta,0)$ is
\bb{red}
\tr_{\mathbb C^4}U_{\epsilon,\vc W}(\beta,0)=1+e^{-\beta\hat\epsilon(\vc p)}\tr_{\mathbb C^2}U_{\epsilon=0,\vc W}(\beta,0)+e^{-2\beta\hat\epsilon(\vc p)}\,.
\ee
Using the H\"older-inequality for traces (see e.g. \cite{14II}), we find that
\bb{Holder}
\big\arrowvert\tr_{\mathbb C^2}\left(U_{0,\vc W}(\beta,0)\right)\big\arrowvert\leq \tr_{\mathbb C^2}\left(U_{0,\vc W=W_0\cdot\vc n}(\beta,0)\right)\,,
\ee
for an arbitrary {\em constant} unit vector $\vc n$.
{F}rom this inequality we can conclude, that $x$-{\em independent} fluctuations of an exchange field of
constant length around a ferromagnetic mean-field configuration do {\em not}
decrease the free energy and hence are not a source of
instability.

Since the calculation is rather easy, we give an explicit formula for the coefficient $C_t$ of the magnetic
$\sigma$-model when the exchange field is ferromagnetically ordered. Because of the stability of the system with respect to
$x$-independent fluctuations we must find a negative sign for
this coefficient. We use the H\"older inequality (\ref{Holder}),
in order to get an upper bound for the large-$\beta$ behaviour of the
trace in (\ref{red}) and find, for large $\beta$, and up to a $\vc W$-independent constant
\bb{Sred}
S_{eff}(\vc W)\approx (V_--V_+)\ln\det(\partial_t+W)+S_K(\vc W)\,,
\ee
where $V_\pm=\operatorname{Vol}\{\vc p|\hat\epsilon(\vc p)\pm W_0/2<0\}$.
In order to calculate $C_t$ 
we take a look at the real part of $\ln\det(\partial_t +W)$,
\begin{align*}
\Re\ln\det\left(\partial_t +W\right)=\frac{1}{2}\ln\det\left[\left(\partial_t +W\right)\left(-\partial_t +W\right)\right]
=\frac{1}{2}\ln\det\left(\Lambda+(\partial_t W)\right)\,,
\end{align*}
where $\Lambda:=-\partial_t^2+(1/4)W_0^2$. We treat $(\partial_t W)$ as a perturbation of $\Lambda$ and proceed in the same way as in the last subsection to arrive at
$$
\Re\ln\det\left(\partial_t + W\right)=-\frac{1}{4}\trs(\Lambda^{-2}(\partial_t W)^2)+\text{h.o.}\,,
$$
where we omitted $\vc W$-independent terms.
Finally, we find for the coefficient $C_t$
\bb{Ct}
C_t=-\frac{W_0^2(V_--V_+)}{8(2\pi)^3\beta}\Sum{k_0}\frac{1}{(k_0^2+W_0^2/4)^2}\overset{\beta W_0\rightarrow\infty}{\longrightarrow}-\frac{V_--V_+}{4 (2\pi)^3 W_0}\,,
\ee
which is indeed manifestly negative.

Next, we investigate the stability of the ferromagnetically ordered mean-field configurations w.r.t. time-independent, but {\em $x$-dependent} perturbations, which amounts to the calculation of $C_{\nabla}$ in the magnetic $\sigma$-model. This calculation is somewhat harder, because the $p$-dependent modes in the operator $D_{\vc W}$ do not decouple.
We perturb the ferromagnetic mean-field configuration $\vc W=W_0\vc n_z$
by time-independent fluctuations, $\pmb{\chi}$, and calculate their
contribution to the real part of the effective action to second
order in $|\pmb{\chi}|/W_0$. Since the effective action has a
quadratic minimum w.r.t. the length of the exchange field, we can
assume $\bchi$ to be orthogonal to $\vc n_z$.
Then (omitting ${\pmb\chi}$-independent terms)
\bb{109a}
\Re S_{eff}(W_0\vc n_z+\pmb{\chi})=-\frac{1}{2}\Sum{x,y}\pmb{\chi}^T(x)K(x-y)\pmb{\chi}(y)-\frac{1}{4}\trs(\Lambda^{-1}\{\epsilon,\chi\}
\Lambda^{-1}\{\epsilon,\chi\})+\text{h.o.}\,,
\ee
where
\begin{align*}
\Lambda &=-\partial_t^2+\epsilon^2+\frac{W_0^2}{4}+W_0\epsilon\sigma_3\,,\\
\chi &=\frac{1}{2}\pmb{\chi}\cdot\pmb{\sigma}\,.
\end{align*}
In the trace we only keep contributions quadratic in $|\vc p|/W_0$, where $\vc p$ is the momentum labelling the modes of $\pmb{\chi}$ (derivative expansion).
Spin degrees of freedom and Matsubara frequencies can be summed over explicitly, and - after a straightforward but tedious calculation - we obtain a formula for the coefficient, $C_{\nabla}$, of the term
$$
\frac{\beta}{W_0^2}\Sum{x}|\nabla\pmb{\chi}(x)|^2\,
$$
in the expansion of the second term on the R.S. of (\ref{109a}),
which, for large values of $\beta$, is given by
\begin{multline}\label{Cnabla}
C_\nabla\approx(2\pi)^{-3}\Int{B_\Gamma}d\vc k\frac{|\nabla\epsilon(\vc k)|^2}{4W_0}\Bigg\{-\Theta\left(\epsilon(\vc k)-\frac{W_0}{2}\right)\frac{\beta W_0}{4}\frac{1}{\cosh\beta\left(\epsilon(\vc k)-\frac{W_0}{2}\right)+1}\\
+\Theta\left(\frac{W_0}{2}-|\epsilon(\vc k)|\right)\left[1-\frac{\beta W_0}{4}\left(\frac{1}{\cosh\beta\left(\epsilon(\vc k)+\frac{W_0}{2}\right)+1}+\frac{1}{\cosh\beta\left(\epsilon(\vc k)-\frac{W_0}{2}\right)+1}\right)\right]\\
-\Theta\left(-\epsilon(\vc k)-\frac{W_0}{2}\right)\frac{\beta W_0}{4}\frac{1}{\cosh\beta\left(\epsilon(\vc k)+\frac{W_0}{2}\right)+1}\Bigg\}\,.
\end{multline}
By plugging $\vc W=W_0\vc n_z+\pmb{\chi}$, with $\pmb{\chi}$
orthogonal to $\vc n_z$, into the $\sigma$-model action (\ref{Fsm}), one
easily sees that the coefficient $C_{\nabla}$ just derived coincides
with the corresponding coefficient in the $\sigma$-model.

For the {\em Hubbard model} we can deduce sufficient conditions for the linear stability of the ferromagnetic mean-field configuration (i.e., for $C_{\nabla}$ to be {\em negative}) from (\ref{Cnabla}), in the regime where $\beta$ is large.
We denote by $v_F^{\pm}$ the Fermi velocities of the upper/lower spin band, namely
$$
|\epsilon(\vc k)\pm \frac{W_0}{2}|\approx v_F^{\pm}|\vc k-\vc
k_F^{\pm}|\,,\quad \text{for $\vc k$ close to the Fermi momenta
$\vc k_F^{\pm}$}\,,
$$
by $A_F^{\pm}$ the area of the Fermi surface of the upper/lower spin
band, and by $V_F^{\pm}$ the respective
volumes. Then, for large $\beta$, a sufficient condition for linear stability is that
\bb{stability}
\frac{W_0}{2}(A_F^+v_F^++A_F^-v_F^-)>v_{\max}^2(V_F^--V_F^+)\,,
\ee
where $v_{\max}=\max\{|\nabla\epsilon(\vc k)|\,|\epsilon(\vc k)-W_0/2<0<\epsilon(\vc k)+W_0/2\}$.

We now combine this stability condition with the mean-field equation (\ref{3.48}) for the Hubbard-model, i.e.,
$$
W_0=\frac{2U_0}{3}\frac{V_F^--V_F^+}{(2\pi)^3}\,,
$$
to arrive at the stability condition
$$
\frac{U_0}{3}\frac{A_F^+v_F^++A_F^-v_F^-}{(2\pi)^3}>v_{\max}^2\,.
$$
Details of these and other related calculations will appear in \cite{G}.

\subsection*{Magnon dynamics and the $WZ$-term}
In this subsection we consider the Wess-Zumino-term in (\ref{Fsm}).
We first consider an $x$-independent exchange field $\vc W(t)$ of constant length $|\vc W(t)|=W_0$ and calculate the variation of the first term on the R.S. of (\ref{3.38}) for $\epsilon=0$, given a variation, $\delta\vc W$, of the exchange field:
\bb{4.55}
\delta\ln\det(D_{0,\vc W})=\trs(\delta D_{0,\vc W} (D_{0,\vc W}^*D_{0,\vc W} )^{-1}D_{0,\vc W} ^*)\,,
\ee
where
$$
\delta D_{0,\vc W}=\frac{1}{2}\delta\vc W(t)\cdot\pmb{\sigma}\equiv \delta W(t)\,.
$$
In order to evaluate the R.S. of (\ref{4.55}), we expand $(D_{0,\vc W}^*D_{0,\vc W})^{-1}$ in a Neumann series in powers of $(\partial_t W)$,
$$
(D_{0,\vc W}^*D_{0,\vc W})^{-1}=\Lambda^{-1}+\Lambda^{-1}(\partial_t W)\Lambda^{-1}+\dots\,,
$$
where $\Lambda=-\partial_t^2+(1/4)W_0^2$.
Plugging this series into (\ref{4.55}), we find that
\bb{4.56}
\delta\ln\det(D_{0,\vc W})=\delta S^{(1)}(\vc W)+\delta S^{(2)}(\vc W)+\dots\,.
\ee
where
\begin{align}
\delta S^{(1)}(\vc W)& =\trs(\delta W\Lambda^{-1}D_{0,\vc W}^*)\,,\label{4.57}\\
\delta S^{(2)}(\vc W)& =\trs(\delta W\Lambda^{-1}(\partial_t W)\Lambda^{-1}D_{0,\vc W}^*)\,,\label{4.58}
\end{align}
etc.. Using the cyclicity of the trace and the fact that $\Tr(\pmb{\sigma})=0$ (where ``$\Tr$'' is the trace on $2\times 2$ matrices), we see that
$$
\delta S^{(1)}(\vc W)=\alpha(W_0)\int\limits_0^\beta dt \Tr(W(t)\delta W(t))\,,
$$
where
$$
\alpha(W_0)=\beta^{-1}\Sum{k_0\in Z_\beta}\frac{1}{k_0^2+(1/4)W_0^2}\,,
$$
and $Z_{\beta}=(\pi/\beta)(2\mathbb Z +1)$. Since
$$
\Tr(W(t)\delta W(t))=\frac{1}{2}\delta\Tr(W(t)^2)\,,
$$
$\delta S^{(1)}$ only depends on variations of the length of $\vc W(t)$ and {\em vanishes} if the length is held fixed.

By (\ref{4.58}),
$$
\delta S^{(2)}(\vc W)=\delta S^{(2)}_I(\vc W)+\delta S^{(2)}_{II}(\vc W)\,,
$$
where
$$
\delta S^{(2)}_I(\vc W)=-\trs(\delta W\Lambda^{-1}(\partial_t W)\Lambda^{-1}\partial_t)=-\frac{1}{2}\sum\limits_{j=1}^3\trs(\delta W_j\Lambda^{-1}(\partial_t W)_j\Lambda^{-1}\partial_t)
$$
and
$$
\delta S^{(2)}_{II}(\vc W)=\trs(\delta W\Lambda^{-1}(\partial_t W)\Lambda^{-1}W)=-\frac{i}{4}\Sum{j,l,m}\epsilon_{jlm}\trs(\delta W_j\Lambda^{-1}(\partial_t W)_l\Lambda^{-1}W_m)\,.
$$
Here we have used that $\Tr(\sigma_j\sigma_l)=2\delta_{jl}$ and
$\Tr(\sigma_j\sigma_l\sigma_m)=-2i\epsilon_{jlm}$. Since
$\partial_t$, $\Lambda$ and $W_j(t)$ are real operators, $\delta
S^{(2)}_I$ is real and $\delta S^{(2)}_{II}$ purely imaginary.
The real part has been calculated in the last subsection.
Turning to $\delta S^{(2)}_{II}(\vc W)$, we observe that
\begin{align}\label{4.63}
\delta S^{(2)}_{II}(\vc W)&=\trs(W\delta W(\partial_t W)\Lambda^{-2})+\text{higher derivative terms}\nonumber\\
     &=-\frac{i}{4}\gamma(W_0)\int\limits_0^{\beta}dt\Sum{j,l,m}\epsilon_{jlm}W_j(t)\delta W_l(t)\partial_tW_m(t)+\text{h.d.t.}\nonumber\\
     &=-\frac{i}{2}\int\limits_0^{\beta}dt\hat{\vc W}(t)\cdot(\delta\hat{\vc W}(t)\wedge\partial_t\hat{\vc W}(t))+\mathcal O((\beta W_0)^{-1})+\text{h.d.t.}\,,
\end{align}
 where, in the last equation, we have used that
\bb{4.61}
\gamma(W_0):=\beta^{-1}\Sum{k_0\in
Z_\beta}\frac{1}{(k_0^2+(1/4)W_0^2)^2}\underset{\beta
W_0\rightarrow\infty}{\longrightarrow}\frac{2}{W_0^3}\,.
\ee
We note that $\text{h.d.t.}\sim\mathcal  O((\beta W_0)^{-1})$.

Recall that $\vc W(t)$ has periodic boundary conditions at $t=0,\beta$. Viewing the imaginary-time circle $[0,\beta)$ as the equator of a sphere $S^2_{\beta}$ of radius $\beta/(2\pi)$, we may extend the unit-vector field $\hat{\vc W}(t)$ from the equator of $S^2_{\beta}$ to a continuous unit-vector field $\hat{\vc W}(t,s)$, $0\leq s\leq \beta/2$, on the entire sphere in an arbitrary way, but with $\hat{\vc W}(t,\beta/4)=\hat{\vc W}(t)$. The first term on the R.S. of (\ref{4.63}) then turns out to be the variation of the functional
\begin{align}
S_{WZ}(\vc W)&=-\frac{i}{2}\int\limits_0^{\beta}dt\int\limits_0^{\beta/4}ds \hat{\vc W}(t,s)\cdot(\partial_s\hat{\vc W}(t,s)\wedge\partial_t\hat{\vc W}(t,s))\nonumber\\
     &=-\frac{i}{2}\Int{S^{2,N}}\hat{\vc W}\cdot(d\hat{\vc W}\wedge d\hat{\vc W})\,,\label{4.64}
\end{align}
where ``$d$'' denotes the exterior derivative, and $S^{2,N}$ is the northern hemisphere of $S^2$. 
We recognize (\ref{4.64}) to be the Wess-Zumino term in (\ref{Fsm}) for an $x$-independent (i.e., ferromagnetically ordered) exchange field. The corresponding coefficient $C_{WZ}$ depends on $\epsilon$ and will be determined later on.
The R.S. of (\ref{4.64}) is expressed in standard differential-form notation, which makes it manifest that $S_{WZ}(\vc W)$ is {\em independent} of the radius, $\beta/2\pi$, of the sphere.

The leading term on the R.S. of (\ref{4.63}) is also the variation of
$$
S'_{WZ}(\vc W)=\frac{i}{2}\Int{S^{2,S}}\hat{\vc W}\cdot(d\hat{\vc W}\wedge d\hat{\vc W})\,,
$$
where $S^{2,S}$ is the southern hemisphere of $S^2$. One observes that
\bb{4.65}
S_{WZ}(\vc W)-S'_{WZ}(\vc W)=-\frac{i}{2}\Int{S^{2}}\hat{\vc W}\cdot(d\hat{\vc W}\wedge d\hat{\vc W})=2\pi in\,,
\ee
where $n=n(\hat{\vc W})\in\mathbb Z$ is the degree of the map $\hat{\vc W}:S^2\rightarrow S^2$. It follows that, apparently, the Wess-Zumino action $S_{WZ}$ is only determined {\em modulo an integer multiple of $2\pi i$}.
Due to (\ref{4.65}), $\exp(S_{WZ}(\vc W))$ is a {\em single-valued} functional of $\vc W$.


If the exchange field $\vc W(x,t)$ is {\em antiferromagnetically} ordered, i.e., the N\'eel field
$$
\vc N(x,t)\equiv(-1)^{|x|}\vc W(x,t)= W_0\vc n\,
$$
is a constant field pointing in the direction of some unit vector $\vc n$, the Wess-Zumino term in (\ref{Fsm}) obviously tends to $0$ in the formal continuum limit. It follows that, in antiferromagnetically ordered
states, at sufficiently low temperatures, there are gapless
Goldstone bosons with dispersion \bb{4.69} \omega(\vc k)\approx
v|\vc k|\,,\quad(|\vc k|\approx 0)\,, \ee for some velocity $v$.


In {\em ferromagnetically ordered systems}, however, the Wess-Zumino term does {\em not} disappear.
Using (\ref{Sred}) and (\ref{4.64}), the coefficient $C_{WZ}$ in (\ref{Fsm})
is found to be
\bb{CWZ}
C_{WZ}=-\frac{i}{2}\frac{(V_--V_+)}{(2\pi)^3}\,,
\ee
in the limit where $\beta W_0\rightarrow\infty$.
Had we {\em normalized} the Fourier modes and were we in a finite volume, we would find that $\exp(\tilde S_{eff}(\vc W))$ is single-valued.
{F}rom (\ref{Fsm}) we derive the equations of motion
\bb{4.70}
C_{WZ}\hat{\vc W}_x\wedge\dot{\hat{\vc W}}_x=\Sum{y}
\tilde K(x-y)\hat{\vc W}_y+\a\hat{\vc W}_x\,,
\ee
 with $\hat{\vc W}_x(t)=\hat{\vc W}(x,t)$, $\dot{\hat{\vc W}}_x=\partial_t\hat{\vc
W}_x$, where $\tilde K(x-y)=W_0^2K(x-y)+C_{\nabla}\delta_{|x-y|,1}$, and where $\a$ is a Lagrange multiplier arising from the constraint $|\hat{\vc W}_x(t)|^2=1$, for all $x$ and all $t$. Taking the vector
product of (\ref{4.70}) with $\hat{\vc W}_x$, and doing the Wick
rotation from imaginary time back to real time \cite{12}, we
obtain the equations of motion for {\em magnons} in a ferromagnet
\bb{4.71}
iC_{WZ}\dot{\hat{\vc W}}_x=\hat{\vc W}_x\wedge\Sum{y}\tilde K(x-y)\hat{\vc W}_y\,,
\ee
which is the well known {\em
Landau-Lifshitz equation}; see also \cite{17}.

If $J$, and hence $K$, are of short range and if the equilibrium state of the system were ferromagnetically ordered, i.e.,
$$
{\vc W}(x,t)=W_0\vc n+\bchi(x,t)\,,\quad W_0>0\,,
$$
where $\vc n$ is a unit vector and $\bchi(x,t)$ is assumed to be small, then the dispersion of magnons is found to be
$$
\omega(\vc k)=\frac{|\vc k|^2}{2M}\,,\quad\text{for some constant $M>0$}\,,
$$
as expected. If $\vc n$ points in the $z$-direction then $\bchi$ lies essentially in the $x-y$ plane.
We define
$$
\phi(x,t)=\chi_1(x,t)+i\chi_2(x,t)\,.
$$
Passing to the formal continuum limit, the action functional for the complex magnon field $\phi$ of a ferromagnet is seen to be given by
\bb{4.72}
S_{\operatorname{magnon}}(\bar{\phi},\phi)=\int\limits_0^{\beta}dt\int dx[\bar{\phi}(x,t)\partial_t\phi(x,t)+\frac{1}{2M}\bar{\phi}(x,t)(\Delta\phi)(x,t)]\,,
\ee
which is the action for a system of {\em conserved non-relativistic
bosons} with vanishing chemical potential. Since $ W_1=\chi_1=(\phi+\bar{\phi})/2$ is an observable
field, the chemical potential actually necessarily vanishes, and there is no
``magnon condensation''; (just like photons do not form any Bose condensates).

\subsection* {Absence of symmetry breaking in one and two dimensions at
  positive temperatures}

For $\beta<\infty$, we can study the fluctuations of the modes $
{\vc W}(k_0,\vc k)$ of the exchange field with $k_0=0$, i.e., of
$\int\limits_0^\beta{\vc W}(x,t)dt$.

We set
$$
\int\limits_0^\beta{\vc W}(x,t)dt=\beta({\vc
  W}^{mf}+\chi_l(x)\vc n_z+\pmb{\chi}_t(x))\,,
$$
where ${\vc W}^{mf} $ is a solution of the ferromagnetic or the antiferromagnetic mean-field equation.
Then, in $d$ spatial dimensions, we conclude from our stability analysis
$$
\langle \pmb{\chi}_t(x)^2\rangle_{\beta,\mu}\propto
\Int{B_\Gamma}\frac{d^dk}{k^2}
$$
which, for $d=1$ and $2$, is infrared-divergent. Thus, transversal
fluctuations around the mean field are gigantic and hence destroy
long-range order; (this is the Mermin-Wagner theorem, see, e.g.,
\cite{B}).

In two dimensions, there is an alternative way to understand the
absence of ordering when $\beta<\infty$: Setting ${\vc
  W}(x,t)=\tilde{\vc W}(x)$ (ferromagnetic short-range order), or ${\vc
  W}(x,t)=(-1)^{|x|}\tilde{\vc W}(x)$ (antiferromagnetic short-range order),
  where $\tilde{\vc W}(x)$ is slowly varying,
and passing to the formal continuum limit (lattice spacing
$\rightarrow 0$), one finds that the action $S_{eff}(\vc W)$ has
(approximate) critical points, indeed local maxima, on
time-independent configurations $\tilde{\vc W}(x)$ with the
properties:
$$
|\tilde{\vc W}(x)| \approx W_0\,,\quad\text{for all}\quad x\,,
$$
where $W_0$ is a solution of the (ferromagnetic or antiferromagnetic)
mean-field equation,
$$
\tilde{\vc W}(x)\rightarrow W_0\vc n\,,\quad\text{as}\quad
|x|\rightarrow \infty\,,
$$
where $\vc n$ is an arbitrary unit vector, and the degree of the map
$\tilde{\vc W}(x) $, as measured by the integer (``winding number'')
$$
\frac{1}{4\pi W_0^3}\int d^2x\tilde{\vc
  W}(x)\cdot\left(\frac{\partial\tilde{\vc W}(x)}{\partial
    x_1}\wedge\frac{\partial\tilde{\vc W}(x)}{\partial x_2}\right)\,,
$$
is non-zero. Such configurations are called ``instantons''. The
contributions of instanton configurations to a functional integral,
such as (\ref{3.41}), destroy long-range order.

{\em Remark.} 
Generally speaking, it is hard to justify the use of steepest descent
in approximately evaluating functional integrals, such as (\ref{3.39})
and (\ref{3.41}), because there isn't any large constant $\mathcal
N=\hbar^{-1}$ multiplying $S_{eff}(\vc W)$. If, however, we place
$\mathcal N$ identical species of spin-$\frac{1}{2}$-fermions on each
site $x\in\Gamma$, coupled to each other only through
exchange interactions between the total spin operators, and if we set
$K_0=\mathcal NK_0$ then $S_{eff}(\vc W)\equiv
S_{eff}^{(K_0)}(\vc W)$ is replaced by $\mathcal N S_{eff}^{(K_0)}(\vc W)$, and steepest descent becomes reliable.

\section{Reflection positivity and phase transitions in Heisenberg-models}
\setcounter{equation}{0}

In order to establish the existence of phase
transitions in some of the models introduced in Section 3, we recall the
method of \emph{reflection positivity}
(\cite{3},\cite{4},\cite{14},\cite{14II}). For simplicity, we choose the
lattice $\Gamma$ to be given by $\bZ^d$. We \emph{decompose}
$\Gamma$ \emph{into two equal halves} \bb{4.11}
\Gamma_+=\{x\in\Gamma|x_1>0 \},\qquad\Gamma_-=\{x\in\Gamma|x_1\leq
0 \},\ee where $x_j$ is the $j^{th}$ component of $x\in\Gamma$.
Let $\Pi$ be the plane in $\bR^d$ orthogonal to the 1-direction at
height  $x_1=\frac{1}{2}$, and let $\vartheta$ denote reflection
at $\Pi$. Clearly, $\vartheta$ maps $\Gamma_-$ onto $\Gamma_+$,
and conversely.

An exchange coupling matrix, $J(x,y)$, is said to be
\emph{reflection positive} iff
\bb{4.12}\sum_{x,y}\overline{f(x)}J(x,\vartheta y)f(y)\geq 0, \ee
for all functions $f$ on $\Gamma$ with $\operatorname{supp} f
\subseteq \Gamma_+$. Note that this condition is
\emph{independent} of the diagonal elements, $J(x,x)$, of $J$; so,
by choosing them appropriately, we can always ensure that $J$ is
either positive-definite or negative-definite. By computing some
integrals with a Gaussian measure of mean 0 and covariance
$-J^{-1}$ ($J$ negative-definite) we see that if $J$ is reflection
positive then so is $-J^{-1}$; see e.g. \cite{14},\cite{14II}. Furthermore, if
$J$ is reflection-positive then so is $-\widetilde{J}$, where
$\widetilde{J}$ is given by \bb{4.13}
\widetilde{J}(x,y)=(-1)^{|x+y|}J(x,y)=(-1)^{|x|+|y|}J(x,y)\ee with
$|x|=\sum_{j=1}^d x_j$. (Note that if $J$ is positive-definite then
$\widetilde{J}$ is positive-definite, too.) The following simple
calculation proves our claim. Let \[
\tilde{f}(x)=(-1)^{|x|}f(x).\] Since $J$ is assumed to be
reflection-positive, \[\begin{array}{rcl} 0 &\leq
&\displaystyle\sum_{x,y}\overline{\tilde{f}(x)}J(x,\vartheta y)\tilde{f}(y)\\
&=&\displaystyle\sum_{x,y}\overline{f(x)}(-1)^{|x|+|y|}J(x,\vartheta y)f(y)\\
&=&\displaystyle
-\sum_{x,y}\overline{f(x)}(-1)^{|x+\vartheta y|}J(x,\vartheta y)f(y)\\
&=&\displaystyle-\sum_{x,y}\overline{f(x)}\widetilde{J}(x,\vartheta
y)f(y),
\end{array}\]
for an arbitrary function $f$ with $\operatorname{supp} f
\subseteq \Gamma_+$ (hence $\operatorname{supp} \tilde{f}
\subseteq \Gamma_+$). These considerations are summarized as
follows: 
\bb{4.14}
\textit{If $J$ is reflection-positive then
$-K:=-J^{-1}$ is reflection-positive, too;} 
\ee 
and \[ \textit{if
$-J$ is reflection-positive (i.e., $J$ is reflection-negative)}\]
\bb{4.15}\textit{then $-\widetilde{K}:=-\widetilde{J^{-1}}$ is
reflection-positive, too.}\ee In both situations, we can choose
the diagonal elements of $J$ such that $J$ and $K$ are positive-,
or negative-definite.

A reflection-positive matrix $J$ is called \emph{ferromagnetic},
while a reflection-negative matrix is called
\emph{antiferromagnetic}.
It is easy to see that, for nearest-neighbour exchange couplings,
these notions of ``ferro-'' and ``antiferromagnetic'' coincide
with the familiar ones.

Let $J$ be ferromagnetic and negative-definite. Then the Gaussian
measure \[ d\mu_{-J}(\vec\phi)=\const\cdot
e^{\frac{1}{2}\sum_{x,y}\vec\phi (x) J(x,y) \vec\phi (y)} \prod_x
d^n \phi(x),\] where $\vec\phi(x)\in\bR^n,\, n=1,2,3,\ldots,$ and
the constant is chosen such that $d\mu_{-J}$ is normalized, is
\emph{reflection-positive}, in the sense that \bb{4.16} \int d\mu
_{-J}(\vec\phi) \overline{(\Theta F)(\vec\phi)}F(\vec\phi)\geq
0,\ee for an arbitrary bounded function $F$ that depends only on the
variables $\{ \vec\phi (x)| x\in\Gamma_+\}$; the space of all such
functions is denoted by $\mathcal{F}_+$. The function $\Theta F$
in (\ref{4.16}) is defined by \bb{4.17}( \Theta
F)(\vec\phi)=F(\vec\phi _\vartheta), \ee where
$\vec\phi_\vartheta(x)=\vec\phi(\vartheta x)$. If
$\vec\phi^{(1)},\ldots,\vec\phi^{(l)}$ are independent Gaussian
random fields with distributions given by
$d\mu_{-J^{(t)}}(\vec\phi^{(t)})$, where $J^{(t)}$ is
ferromagnetic and $J^{(t)}<0,\, t=1,\ldots,l$, then the
distribution \bb{4.18} d\mu(\vec\phi)=\prod_{t=1}^l
d\mu_{-\tau^{(t)}J^{(t)}}(\vec\phi ^{(t)})\ee of the random field
$\vec\phi$ given by \[ \vec\phi(x,t)=\vec\phi^{(t)}(x),\,
t=1,\ldots,l,\] is reflection-positive, for arbitrary
$\tau^{(t)}>0,\,t=1,\ldots,l$; as is easily seen.

These considerations have the following consequence that will play
a crucial role in our analysis of phase transitions: Let $\vc
W(x,t),\,x\in\Gamma,\,t\in [0,\b)$, with $\vc W(x,t+\b)=\vc
W(x,t)$, be the  \emph{imaginary time exchange field}, and let
$d\mu_K(\vc W)=\exp(S_K(\vc W))\mathcal{D}\vc W$ be its Gaussian
distribution, as introduced in Section \ref{sec:mf}, (\ref{3.25})
through (\ref{3.29}). If the exchange couplings $J$ are {\it
ferromagnetic} (in the sense specified above) and
positive-definite then $-K=-J^{-1}$ is ferromagnetic and
negative-definite; hence the Gaussian measure \bb{4.19} d\mu_K(\vc
W)\, \textrm{ is reflection-positive,}\ee as follows from (\ref{3.27}),
(\ref{4.16}) and (\ref{4.18}).

Next, we suppose that the exchange couplings $J$ are {\it
antiferromagnetic} (in the sense specified above). Then, by
(\ref{4.15}), $-\widetilde{K}=-\widetilde{J^{-1}}$ is {\it
reflection-positive} and negative-definite, for $J$
positive-definite. We introduce a random field, $\vc N$, (``N''
for ``N\'eel''), by setting \bb{4.20} \vc N(x,t)=(-1)^{|x|}\vc
W(x,t). \ee If $\vc W$ has distribution $d\mu_K(\vc W)$, with
$K=J^{-1}$, then $\vc N$ has distribution
$d\mu_{\widetilde{K}}(\vc N)$, and hence \bb{4.21}
d\mu_{\widetilde{K}}(\vc N)\, \textrm{ is reflection-positive.}\ee
We set \bb{4.22} \vc N_\vartheta(x,t)=\vc N(\vartheta x,t),\ee and
(see (\ref{4.17})) \bb{4.23} (\Theta F) (\vc N)=F(\vc
N_\vartheta),\ee for an arbitrary function $F$ of $\{\vc
N(x,t)|x\in\Gamma,\,t\in[0,\b)\}$.

We consider the models introduced in Section \ref{sec:mf} without hopping term but with $\mathcal N$ Fermion species.
They have an action functional given by
\begin{multline}\label{4.1}
\bar S(\bar C,C;\vc W)=\sum\limits_{a=1}^{\mathcal N}\int\limits_0^{\beta}dt\Sum{x}\left[\bar C^{(a)}_s(x,t)\left(\frac{\partial}{\partial t}-\mu\right)C^{(a)}_s(x,t)-\nu_a\vc W(x,t)\cdot\vc S^{(a)}(x,t)\right] \\
-\frac{1}{2}\int\limits_0^{\beta}dt\left(\Sum{x,y}\vc W(x,t)K(x-y)\vc W(y,t)\right)\,,
\end{multline}
where $\nu_a=\pm 1$, for all $a=1,\dots,\mathcal N$, and $\hat
K(k)=\left(\frac{4}{3}U_0+\hat J_{\neq}(k)\right)^{-1}$; see
(\ref{3.19}), (\ref{3.27}), and (\ref{3.30}) through (\ref{3.32}).
Since the first term on the R.S. of (\ref{4.1}) does not couple
different sites in the lattice, the effective action of the
exchange field $\vc W$, see (\ref{3.34}) and (\ref{3.38}), is
given by \bb{4.2} S_{eff}(\vc
W)=-\frac{1}{2}\int\limits_0^{\beta}dt\left(\Sum{x,y}\vc
W(x,t)K(x-y)\vc W(y,t)\right)+\Sum{a}\Sum{x}\ln\det(D_{\nu_a\vc
W(x,\cdot)})\,, \ee with $D_{\vc W}$ given in (\ref{3.36}), for
$\hat t(x)\equiv 0$. We define \bb{4.24}
\begin{array}{rcl}
f_x(\vc N)&=&\displaystyle\prod_{a=1}^\mathcal{N} \det(D_{\nu_a(-1)^{|x|}\vc N(x,\cdot)})\\
&=&\displaystyle\prod_{a=1}^\mathcal{N} \det(D_{\nu_a\vc
W(x,\cdot)}),
\end{array}
\ee
Then it follows that \bb{4.25}
\begin{array}{rcl}
\overline{\Theta f_x(\vc N)} &=&\displaystyle
\overline{f_x\big(\vc N
_\vartheta (x,\cdot)\big)}\\
&=&\displaystyle\prod_{a=1}^\mathcal{N}\overline{\det(
D_{\nu_a(-1)^{|x|}\vc
N _\vartheta (x,\cdot)})}\\
&=&\displaystyle\prod_{a=1}^\mathcal{N}\overline{\det(
D_{-\nu_a(-1)^{|\vartheta x|}\vc N (\vartheta x,\cdot)})}\\
&=&\displaystyle \prod_{a=1}^\mathcal{N} \det(
D_{\nu_a(-1)^{|\vartheta x|}\vc
N (\vartheta x,\cdot)})\\
&=& f_{\vartheta x}(\vc N),
\end{array}\ee
where, in the third equation, we have used that
$(-1)^{|x|}=-(-1)^{|\vartheta x|}$, and, in the fourth equation,
we have inserted identity (\ref{4.10}). Equation (\ref{4.25})
implies that \bb{4.26} \overline{\Theta \Big(
\prod_{x\in\Gamma_+}f_x(\vc N) \Big)}=\prod_{x\in\Gamma_-}f_x(\vc
N).\ee {F}rom this equation and equations (\ref{4.21}) and
(\ref{4.16}) we conclude the following simple, but quite
fundamental
\\ \, \\
\textbf{Result A} Suppose the exchange couplings $J$ are
\emph{antiferromagnetic} (in the sense introduced above) and
positive-definite. We set
\begin{align}\label{4.27}
d\mu_{eff}(\vc N)&=\const\cdot\Prod{x}f_x(\vc N)d\mu_{\widetilde{K}}(\vc N)\\
&=\const'\cdot\exp\Big( S_{eff}(\vc N)\Big)\mathcal{D}\vc N,
\end{align}
where the constants are chosen such that $\int\! d\mu_{eff}(\vc
N)=1$, and
\begin{equation}\begin{split}\label{4.28} S_{eff}(\vc
N) := \; & -\frac{1}{2}\int\limits_0^b dt \left(\Sum{x,y}\vc
N(x,t)\widetilde{K}(x-y)\vc N(y,t)\right)\\  &+
\Sum{a}\Sum{x}\ln\det\Big( D_{\nu_a(-1)^{|x|}\vc N(x,\cdot)}\Big)\,,
\end{split}
\end{equation}
with $\widetilde{K}=\widetilde{J^{-1}}$; see (\ref{4.2}), (\ref{4.20}), (\ref{4.21}).
\\ \, \\
Then the measure $d\mu_{eff}(\vc N)$ is
\emph{reflection-positive}, in the sense that \bb{4.29}
\int\!d\mu_{eff}(\vc N)\overline{(\Theta F)(\vc N)}F(\vc N)\geq
0,\ee for arbitrary functions $F\in\mathcal{F}_+$ (where
$\mathcal{F}_+$ is defined right above (\ref{4.17})).\\ \, \\

Next, we suppose that the number of fermion species,
$\mathcal{N}=2\mathcal{M}$ is \emph{even}, with
\bb{4.30}
\nu_1=\ldots=\nu_\mathcal{M}=1,\,\nu_{\mathcal{M}+1}=\ldots=\nu_{2\mathcal{M}}=-1.
\ee
We define \[ g_x(\vc W)=\det(D_{\vc
W(x,\cdot)})^\mathcal{M}\det(D_{-\vc W(x,\cdot)})^\mathcal{M}.\]
Identity (\ref{4.10}) then implies that $g_x(\vc W)\geq 0$, for
all $x\in\Gamma$, and, setting \[ \vc W _\vartheta (x,t)=\vc W
(\vartheta x,t),\] we find that
\bb{4.31}
\overline{\Theta g_x(\vc
W)}=\Theta g_x(\vc W)=g_{\vartheta x} (\vc W).
\ee
Defining
\begin{align}\label{4.32}
d\mu_{eff}(\vc W)&=\const\cdot\Prod{x} g_x(\vc W) d\mu_K (\vc W)\\ 
&=\const'\cdot\exp\Big( S_{eff}(\vc W)\Big)\mathcal{D}\vc W,
\end{align}
where $K=J^{-1}$, $S_{eff}(\vc W)$ is as in (\ref{4.2}), and the
constants are chosen such that $\int\!d\mu_{eff}(\vc W)=1$, we
arrive at
\\ \, \\ \textbf{Result F} Suppose the exchange couplings
$J$ are \emph{ferromagnetic} (in the sense introduced above). Then
the measure $d\mu_{eff}(\vc W)$ is \emph{reflection-positive},
i.e. \bb{4.33} \int\!d\mu_{eff}(\vc W)\overline{(\Theta F)(\vc
W)}F(\vc W)\geq 0,\ee for arbitrary functions
$F\in\mathcal{F}_+$.\\ \, \\

{F}rom Results A and F one obtains the following \emph{infrared
(spin-wave) bounds}, originally discovered in \cite{3} and
generalized in \cite{14},\cite{14II}.
\\ \, \\
{\em (IRA)} Under the hypotheses of Result A (in particular, $J$
\emph{antiferromagnetic}), one has the inequalities \bb{4.34}
\begin{array}{rcl}
0 & \leq & \langle \widehat{\vc N}(\vc k,k_0)\cdot\widehat{\vc N}(-\vc k,-k_0)\rangle_{\b,\mu} \\
 &\leq & \b W_0^2\delta_0(\vc k)\delta_{k_0,0}+\Big( \widehat{\widetilde{K}}(\vc k)-\widehat{\widetilde{K}}(0)\Big)^{-1}
\end{array}\ee
for some constant $W_0^2\geq 0$. Note that $\widehat{\vc
N}(-\vc k,-k_0)=\overline{\widehat{\vc N}(\vc k,k_0)}$, because $\vc
N(x,t)$ is real.

If $W_0^2$ is strictly positive then the state $\langle
(\cdot)\rangle_{\b,\mu}$ exhibits \emph{long-range order}, and if
we couple $\vc N$ to an arbitrarily small external field, $\e\vc
n$, $|\vc n|=1$, then
\bb{4.35}
\lim_{\e\to 0} \langle \vc N (x,t)
\rangle_{\b,\mu,\e \vc n} = W_0\vc n.\,\,\footnote{This claim is heuristic, but can be replaced by an equivalent, mathematically rigorous one, \cite{14}.}
\ee
Inequality (\ref{4.34})
implies that, for an arbitrary bounded function $f(x,t)$ of rapid
decay in $x\in\Gamma$ and periodic in $t\in[0,\b)$,
\bb{4.36}
\begin{array}{rcl} 0 & \leq & \langle | \vc N(f)|^2
\rangle_{\b,\mu}\\
& \leq & \displaystyle \b W_0^2
|\hat{f}(0,0)|^2+\sum_{k_0\in\frac{2\pi}{\b}\bZ}\,\,\Int{B_\Gamma}\!d\vc k
\frac{|\hat{f}(\vc k,k_0)|^2}{\widehat{\widetilde{K}}(\vc
k)-\widehat{\widetilde{K}}(0)}\,,
\end{array}
\ee
where $\vc N(f)=\int\limits_0^\beta dt\Sum{x}\vc N(x,t)f(x,t)$.

Returning to identities (\ref{3.40}) and (\ref{3.41}), with
\bb{4.37}
\vc S(x,t)=\sum_{a=1}^\mathcal{N}\nu_a \vc S^{(a)}(x,t),
\ee
and recalling that $\vc W(x,t)=(-1)^{|x|}\vc N(x,t)$, we
conclude from (\ref{4.36}) that
\begin{align}\label{4.38}
0 &\leq  \langle(\int\limits_0^\b\!\vc S(0,t)dt)^2
\rangle_{\b,\mu}\\
& \leq  \displaystyle \b \tK(0) +\b W_0^2
\widehat{\tK}(0)^2 +\Int{B_\Gamma}\!d\vc k
\frac{\widehat\tK(\vc k)^2}{\widehat{\widetilde{K}}(\vc k)-\widehat{\widetilde{K}}(0)}\,.
\end{align}
We recall that $-\tK$ is reflection-positive, and the diagonal
elements are chosen such that $\tK$ (or, equivalently, $K$ and
$J$) is positive-definite. Therefore $\widehat\tK(\vc k)>0$, for all
$\vc k\in B_\Gamma$. Furthermore, if $J$ is of short range then $\tK$
is of short range, too, i.e., $\tK(x)$ decreases rapidly in $|x|$,
and, by reflection-positivity of $-\tK$,
$\widehat\tK(0)\leq\widehat\tK(\vc k)$. It follows that
\bb{4.39}
0\leq
\widehat\tK(\vc k)-\widehat\tK(0)\leq \kappa |\vc k|^2,
\ee
for some
constant $\kappa$. Thus, under the assumptions just specified, the
third term on the R.H.S. of (\ref{4.38}) is bounded above by a
\emph{finite} constant, $I(K,d)$, in dimension $d\geq 3$, but is
\emph{(infrared-) divergent} in one and two dimensions;
($I(K,d=1,2)$ is finite \emph{only} if $J$ is of
\emph{long-range}; see \cite{14},\cite{14II}).

A fairly elementary inequality due to Bruch and Falk, see
\cite{4}, yields the following lower bound 
\bb{4.40}
\langle(\int\limits_0^\b\!\vc S(0,t)dt)^2 \rangle_{\b,\mu}\geq
C(\b,\mu,\mathcal{N})\b \ee for some constant
$C(\b,\mu,\mathcal{N})\geq 0$. If the on-site repulsion $U_0$, the
chemical potential $\mu$ and the value of
$\tK^{-1}(0)=\int_{B_\Gamma}\!d\vc k \widehat{\tK}{}^{-1}(\vc k)$ are tuned
appropriately such that, at zero temperature,  the band of each
species of fermions is half-filled then $C(\b,\mu,\mathcal{N})>0$
and $\lim_{\mathcal{N}\to\infty} C(\b,\mu,\mathcal{N})=\infty$.

Thus, we conclude that, for appropriate choices of $U_0$, $\mu$
and $\tK^{-1}(0)$, and for a \emph{sufficiently large number},
$\mathcal{N}$, of fermion species, \bb{4.41} W_0 \textit{ is
strictly positive, for $\beta$ large enough}\ee

Hence, by (\ref{3.40}) and (\ref{4.35}), \bb{4.42}
\begin{array}{rcl}
\displaystyle \lim_{\e\to 0}\langle \vc S (x,0)
\rangle_{\b,\mu,\e\vc n} &=& \displaystyle-\sum_y
K(x-y)\lim_{\e\to 0}\langle \vc
W(y,0)\rangle_{\b,\mu,\e\vc n}\\
&=&\displaystyle-(-1)^{|x|}\sum_y \tK(x-y)\lim_{\e\to 0}\langle
\vc
N(y,0)\rangle_{\b,\mu,\e\vc n}\\
&=&\displaystyle-(-1)^{|x|}\widehat\tK(0) W_0\vc n \neq 0,
\end{array}\ee
with $M=\widehat\tK(0) |W_0|$ the local magnetization, i.e., the
system exhibits \emph{N\'eel order}, for sufficiently large values
of $\b$ and $\mathcal{N}$. A similar result has first been proven
in \cite{4}.

Arguments related to those used in the proof of the Mermin-Wagner
theorem (see \cite{B}) show that if $M>0$ then there are gapless spin waves (see e.g. \cite{15}), in accordance with the mean-field
picture of Section \ref{sec:4}.

Next, we exploit Result F. It implies that, under condition
(\ref{4.30}), the following infrared bounds hold.
\\ \, \\
{\em (IRF)} Under the hypotheses of Result F (in particular, $J$
\emph{ferromagnetic} and (\ref{4.30}) holds), the inequalities
\bb{4.43}\begin{array}{rcl} 0 & \leq & \langle \widehat{\vc
W}(\vc k,k_0)\cdot\widehat{\vc
W}(-\vc k,-k_0)\rangle_{\b,\mu}\\
&\leq & \b W_0^2 \delta_0(\vc k)\delta_{k_0,0}+ \Big(
\widehat{K}(\vc k)-\widehat{K}(0) \Big)^{-1}
\end{array}\ee
for a constant $W_0^2\geq 0$, are valid. (See \cite{3},\cite{14},\cite{14II}). The consequences of these inequalities are
perfectly analogous to those discussed for antiferromagnets
above. We set \bb{4.44} \vc S(x,t) = \sum_{a=1}^\mathcal{M}\Big(
\vc S^{(a)}(x,t)-\vc S^{(a+\mathcal{M})}(x,t) \Big),\ee see
(\ref{4.30}). Then, under appropriate conditions on $U_0$, $\mu$
and $K^{-1}(0)=J(0)$, and for $d\geq 3$,
\bb{4.45} M=\lim_{\e\to 0}\langle\vc
S(x,t)\rangle_{\b,\mu,\e\vc n}=-\widehat{K}(0)W_0\vc n\neq 0
\ee
for $\b$ and $\mathcal{M}$ large enough, which proves long-range
order at low temperatures, for $\mathcal{M}$ large enough;
(presumably $\mathcal{N}=2\mathcal{M}=2$ suffices if $J$ is nearest-neighbour and $d\geq 3$).

Unfortunately, there is, as yet, no such result on the existence
of a phase transition in a quantum Heisenberg \emph{ferromagnet}
with only one species of fermions (electrons), i.e.,
$\mathcal{N}=1$. The reason is that this model is \emph{not}
described by a reflection-positive distribution, $d\mu_{eff}(\vc
W)$, for the exchange field $\vc W$. Setting \bb{4.46} f_x(\vc
W)=\det(D_{\vc W(x,\cdot)}),\ee $d\mu_{eff}$ is given by \bb{4.47}
d\mu_{eff}(\vc W)=\const\cdot \prod_x f_x(\vc W) d\mu_K(\vc W).\ee
Now, if $J$ is ferromagnetic $d\mu_K(\vc W)$ is
reflection-positive if we set $ \vc W_\vartheta (x,t)=\vc W
(\vartheta x,t)$. However, using (\ref{4.10}),
\[\begin{array}{rcl}
\overline{\Theta f_x(\vc W)} &=& \overline{\det(D_{\vc
W(\vartheta x,\cdot)})}\\
&=& \det(D_{-\vc W(\vartheta x,\cdot)})\\
&=& f_{\vartheta x}(-\vc W)\neq f_{\vartheta x}(\vc W).
\end{array}\]
Thus, $d\mu_{eff}(\vc W)$, as given in (\ref{4.47}), is \emph{not}
reflection-positive, and there is therefore \emph{no} reason why
the Infrared Bounds (IRF) should hold. A very similar problem is
encountered in the study of \emph{Bose-Einstein condensation} in
lattice models of non-relativistic, interacting bosons.

In order to get some preliminary insights into phase transitions
for the Heisenberg \emph{ferromagnet}, one may consider the
following \emph{``static approximation''}: one replaces $f_x(\vc
W)$ by \bb{4.48} f_x^{(0)}(\vc W)=\det(D_{\vc W^{(0)}(x)}),\ee
where
\[\vc W^{(0)}(x)=\frac{1}{\b}\int_0^\b\!dt\,\vc W(x,t)\quad\Big(
=\frac{1}{\b}\vc W(x,k_0=0) \Big),
\] which is time-independent. Then
\bb{4.49} \overline{\Theta f_x^{(0)}(\vc W)}=f^{(0)}_{\vartheta x}
(-\vc W)=f^{(0)}_{\vartheta x} (\vc W),\ee as is easily checked, and
$f_x^{(0)}(\vc W)>0$. Replacing $d\mu_{eff}(\vc W)$ by
\[d\mu_{eff}^{(0)}(\vc W)=\prod_x f_x^{(0)}(\vc W)d\mu_K(\vc W),
\]
we conclude that $d\mu_{eff}^{(0)}(\vc W)$ is reflection-positive
if $J$ is ferromagnetic, hence the infrared bounds (\ref{4.43})
hold, and we conclude that, in the \emph{static approximation},
for appropriate choices of $U_0$, $\mu$ and $J(0)$, a phase
transition accompanied by continuous symmetry breaking occurs at
sufficiently low temperatures.

If one replaces electrons with spin $\frac{1}{2}$ by fermions with
spin $s$, or by $2s$ identical species of spin-$\frac{1}{2}$
fermions (with $\nu_a=1$, for all $a=1,\ldots,2s$), and if one
then takes $s\to\infty$ (rescaling the spin operators, $\vc S(x,t)
\mapsto \frac{1}{s} \vc S(x,t)$) then the static approximation
becomes \emph{exact}. This is not surprising, because the limit
$s\to\infty$ corresponds to the \emph{classical limit}, as shown
in \cite{16}, and the spin-$s$ Heisenberg model approaches the
classical Heisenberg model for which the existence of a phase
transition has been established in \cite{3}. Our formalism,
involving the exchange field $\vc W$, offers a neat and simple way
of recovering some of the results in \cite{16}; (we leave this as an
exercise to the reader).
\\ \, \\
{\em Remarks.}
\begin{enumerate}
\item We have outlined a proof of existence of a phase transition
for a class of Heisenberg \emph{antiferromagnets}; (the proof is
mathematically rigorous; some missing details can be inferred from
\cite{3}, \cite{4}). Of course, this is not a surprise. In
\cite{4}, Dyson, Lieb and Simon have already proven a similar
result \emph{also} using reflection positivity (albeit in a
different manner). Their method of proof, too, breaks down for
quantum \emph{ferromagnets}.

\item It appears that to understand phase transitions in quantum
\emph{ferromagnets} and the related phenomenon of
\emph{Bose-Einstein condensation} for lattice gases of
non-relativistic bosons in a \emph{mathematically rigorous way},
we would have to resort to a full-fledged renormalization group
analysis - a technically rather demanding task.

\item The fact that the function $f_x(\vc W)$ defined in
(\ref{4.46}) (see also (\ref{4.27})) is neither positive, nor even
real is the origin of the \emph{``sign (complex phase) problem''}
in numerical simulations of quantum ferro- \emph{and}
antiferromagnets involving, e.g., the Monte Carlo method applied
to (\ref{4.47}). The great quality of methods based on reflection
positivity is that they are applicable to the analysis of certain
\emph{complex} distributions $d\mu_{eff}$.

\item If one considers models in $d$ dimensions with $SO(2n)$ spins, $n=1,2,3,\dots$, then for $d\geq 3$ one can prove a ferromagnetic phase transition, using arguments very similar to those described above.

\item After completion of this paper, work of Bach et al. \cite{Bach} appeared to which we draw the reader's attention.
\end{enumerate}

\label{sec:reflpos}


\end{document}